\documentclass[12pt]{article}
\usepackage{color}
\usepackage{amssymb}
\usepackage{amsmath}
\usepackage[dvips]{graphicx}
\usepackage{mathrsfs}
\usepackage[mathscr]{eucal}
\usepackage{caption}
\usepackage{subcaption}
\newcommand{\bs}{\boldsymbol}
\author{Mariusz Adamski$^1$, Janusz J{\c{e}}drzejewski$^1$ and Taras Krokhmalskii$^2$\\
$^1$Institute of Theoretical Physics, University of Wroc\l aw,\\
pl. Maksa Borna 9, 50--204 Wroc\l aw, Poland\\
$^2$Institute for Condensed Matter Physics,\\
1 Svientsitski Street, 79011 Lviv, Ukraine}

\title{Quantum phase transitions and ground-state correlations in BCS-like models}

\begin{document}
\maketitle
\tableofcontents
\begin{abstract}
\noindent
We study ground-state correlation functions in one- and two-dimensional lattice models of interacting spinful  fermions -- BCS-like models, which exhibit continuous quantum phase transitions. The considered models originate from a two-dimensional model of d-wave superconductivity proposed by Sachdev.
Due to the exact diagonalizability of the considered models in any dimensionality, exact phase diagrams, with several kinds of quantum-critical points, are constructed and closed-form analytic expressions for two-point correlation functions are obtained.
In one- and two-dimensional cases we provide analytic expressions for the asymptotic behavior of those correlation functions at large distances and in neighborhoods of quantum-critical points. The novelty of our results is that in two-dimensions explicit expressions for direction-dependent correlation lengths in terms of model parameters and the values of direction-dependent universal critical indices $\nu$, that characterize the divergence of correlation lengths on approaching critical points, are determined. Moreover, specific scaling properties of correlation functions with respect to parameters of underlying Hamiltonians are revealed.
Besides enriching the knowledge of properties of lattice fermion systems exhibiting continuous quantum phase transitions, especially in two dimensions, our results open new possibilities of testing unconventional methods of studying quantum phase transitions, as the promising fidelity approach or the entanglement approach, beyond one-dimension and beyond the realm of paradigmatic XY and Ising chains in transverse magnetic fields.
\end{abstract}

\section{\label{intro} Introduction}

In recent years, quantum phase transitions and quantum critical phenomena constitute a subject of great interest and vigorous studies in condensed matter physics. Both, experimental and theoretical developments point  out to the crucial role that quantum phase transitions play in physics of frequently studied high-$T_c$ superconductors, rare-earth magnetic systems, heavy-fermion systems or two-dimensional electrons liquids exhibiting fractional quantum Hall effect \cite{sachdev QPTs}, \cite{vojta QPTs}. Quantum critical phenomena have been also observed  in exotic systems as magnetic quasicrystals \cite{deguchi} and in artificial systems of ultracold atoms in optical lattices \cite{zhang}. The so called classical, thermal phase transitions originate from thermal fluctuations, a competition of internal energy and entropy, and are mathematically manifested as singularities in temperature and other thermodynamic parameters of various thermodynamic functions, and such characteristics of correlation functions as the correlation length, at nonzero temperatures. In contrast, quantum phase transitions originate from purely quantum fluctuations and are mathematically manifested as singularities in system parameters of the ground-state energy density, which is also  the zero-temperature limit of the internal energy density. Naturally, singularities of thermodynamic functions appear only in the thermodynamic limit. The importance of quantum phase transitions for physics and the related wide interest in such  transitions stems from the fact that, while a quantum phase transition is exhibited by ground states, hence often termed a zero-temperature phenomenon, its existence in a system exerts a great impact on the behavior of that system also at nonzero temperatures, in some case at unexpectedly high temperatures \cite{kopp chakravarty}, \cite{kinross}.

Theoretically, quantum phase transitions can be studied in quite complex quantum systems by qualitative and approximate methods, or in relatively simple but exactly solvable models by means of analytic methods and high-accuracy numerical calculations \cite{sachdev QPTs}. Naturally, for the purpose of testing and illustrating general or new ideas the second route is most suitable. Traditionally, this route involves studying the eigenvalue problem of a Hamiltonian, the ground state and excitation gaps, determining quantum critical points and symmetries, constructing local-order parameters, calculating two-point correlation functions and their asymptotic behavior at large distances and in vicinities of quantum critical points, with correlation lengths and the universal critical indices $\nu$ that characterize the divergence of correlation lengths on approaching critical points. Carrying out such a programme is a hard task, which has been accomplished only in a few one-dimensional models.
Among those models, there are quantum spin chains as the isotropic and anisotropic XY models in an external transverse magnetic field, including their extremely anisotropic version - the Ising model \cite{sachdev QPTs}. Only in one dimension those models are equivalent to lattice gases of spinless fermions, which can exactly be diagonalized, and exact results concerning the phase diagram, quantum critical points, correlation functions and dynamics have been obtained (concerning XY model see \cite{lieb}, \cite{barouch I}, \cite{barouch II},  concerning the Ising model see \cite{pfeuty}, and for both models \cite{parkinson}). Needless to say that parallel results for a higher-dimensional model are desirable; this is the first motivation of our investigations presented in this paper.

In the last decade, fresh ideas coming from quantum-information science entered the field of quantum phase transitions. One of them is the hypothesis, the so called fidelity approach. It claims that it is possible to locate critical points \cite{zanardi paunkovic 06}, \cite{barjaktarevic 08}, \cite{zhou 08}
and to determine the correlation lengths and universal critical indices $\nu$ by studying (typically numerically) scaling properties,
with respect to the size of the system and the parameters of the underlying Hamiltonian, of the so called quantum fidelity of two ground states in a vicinity of a critical point \cite{rams PRL 11}, \cite{rams PRA 11}. To extract the index $\nu$, the scaling laws of quantum fidelity, derived by renormalization group arguments \cite{venuti zanardi 07}, \cite{albuquerque 10}, are needed.
The task of determining scaling properties of the quantum fidelity should be much easier than the task of calculating large-distance behavior of two-point correlation functions. The most comprehensive results concerning the verification of the fidelity approach have been obtained for one-dimensional quantum spin systems in a perpendicular magnetic field \cite{rams PRL 11} (the case of Ising model), \cite{rams PRA 11} (the case of XY model). These results are very promising: except a vicinity of a multicritical point, the fidelity approach works fine. In order to verify the effectiveness of this approach in dimensions higher than one, we need an at least two-dimensional exactly solvable model, whose quantum critical points, correlation lengths and critical indices in their vicinities are known; this is the second motivation of our investigations reported in this paper.

To go beyond the one-dimensional case, we consider lattice fermion models which originate from the two-dimensional model of d-wave superconductivity proposed by Sachdev \cite{sachdev 02}(see also \cite{sachdev QPTs}), which are spinful BCS-like models. General, mathematical considerations of some classes of such models, but without specifying hopping intensities or coupling constants, which therefore do not reach such subtleties as quantum critical points or critical behavior of correlation functions, can be found in \cite{zanardi 07}, \cite{cozzini}. For translation-invariant hopping intensities and coupling constants the considered models are exactly diagonalizable in any dimension. Consequently, it is possible to derive analytical formulae for correlation functions of finite systems and in the thermodynamic limit. To limit further the great variety of possible models, we restrict the hopping intensities to nearest neighbors and dimensionality $d{\leq}2$.
For $d{=}2$, we choose the underlying lattice as a square one and require the hopping intensities to be invariant under rotations by $\pi/2$. Similarly, we require that the interactions of our systems do not extend beyond nearest neighbors and for $d{=}2$ they are either invariant under rotations by $\pi/2$ (the symmetric case) or change sign after such a rotation (the antisymmetric case). In this way, we end up with a unique model in one dimension and with only two models in two dimensions.

The general plan of the paper is as follows. In section \ref{models} we  define the models considered in the paper and give closed-form formulae for two basic two-point correlation functions. In sections and subsections that follow we limit our considerations to one of those correlation functions -- an offdiagonal matrix element of the ground-state one-body reduced density operator. For a fixed lattice direction, it depends on the distance and the parameters of the underlying Hamiltonian. We focus on the behavior of that function in three regions of its variables: as the distance grows indefinitely, as parameters of the Hamiltonian approach a quantum critical point, and as besides the distance being sufficiently large the parameters approach a quantum critical point. In particular, we determine the correlation lengths and the critical indices $\nu$, providing analytical and numerical results, which are in excellent agreement.

For completeness, in section \ref{1d} we present results obtained for the unique one-dimensional model.
Some of the results of section \ref{1d} have already been used in \cite{ajk-2} to verify the effectiveness of the fidelity approach in the one-dimensional version of our models, similarly to \cite{rams PRL 11} and \cite{rams PRA 11}. Then, section \ref{2d} is devoted to the two-dimensional models, the symmetric and the antisymmetric ones. As compared to most frequently considered one-dimensional models -- spin chains, the novel feature of two-dimensional systems is that the correlation functions depend on lattice directions. We present their asymptotic behavior separately for the diagonal direction and for offdiagonal directions. As a result, section \ref{2d} splits into four independent subsections, each one referring to a different case specified by the model and lattice direction. To facilitate finding a particular result, the contents of each subsection and the order of presenting the results parallel that of section \ref{1d}.
For more detailed comments concerning the contents of subsections see the last two paragraphs of next section.
In section \ref{scaling} we give a resume of numerous scaling laws derived for the two-point correlation function and its correlation length in all the considered models. Some general observations concerning the correlation length are included in this section. Finally, in section \ref{summ}, we formulate again our motivations, summarize our results, pointing out those that we consider most important.

\section{\label{models} The models, their ground states and ground-state correlation functions}

We consider a $d$-dimensional spinful fermion model, given by the Hamiltonian,
\begin{equation}
H =  \sum_{{\bs l},i,\sigma} \left[ \frac{t}{2}
\left(a^{\dagger}_{{\bs l},\sigma} a_{{\bs l}+{\bs e}_i,\sigma} + \mathrm{h.c.} \right)
- \frac{\mu}{d} a^{\dagger}_{{\bs l},\sigma} a_{{\bs l},\sigma} \right]
- \frac{J}{2} \sum_{{\bs l},i} \left[ \Delta_i  \left(
a^{\dagger}_{{\bs l},\uparrow}a^{\dagger}_{{\bs l}+{\bs e}_i,\downarrow} -
a^{\dagger}_{{\bs l},\downarrow}a^{\dagger}_{{\bs l}+{\bs e}_i,\uparrow} \right)
+ \mathrm{h.c}\right],
\label{ham1}
\end{equation}
where $a^{\dagger}_{{\bs l},\sigma}$, $a_{{\bs l},\sigma}$ stand for creation and annihilation operators, respectively, of a spin $1/2$ fermion, with spin projection on a chosen axis $\sigma {=} \uparrow,\downarrow $, in a state localized at site ${\bs l}{=}(l_1,\ldots,l_d)$ of a $d$-dimensional hypercubic lattice. The edge of the lattice in the direction given by the unit vector ${\bs e}_i$, $i{=}1,\ldots,d$, whose $m$-th component is $\delta_{i,m}$, consists of $L_i$ equidistant sites, labeled by $l_i{=}0,1,\ldots,L_i-1$, $l_j{=}0$ for $j{\neq}i$. In all the considerations that refer to finite systems, special boundary conditions, specified below, are chosen.
The sums over ${\bs l},i$ in (\ref{ham1}) amount to the sum over pairs of nearest neighbors, with each pair counted once. The real and positive parameter $t$ is the nearest-neighbor hopping intensity, $\mu$ -- the chemical potential, $J$ -- the coupling constant  of the gauge-symmetry breaking interaction, and $\Delta_i$, $i{=}1,\ldots,d$, stand for direction-dependent, in general complex, dimensionless constants. Naturally, we can express the parameters $\mu$ and $J$ in units of $t$, while the lengths of the underlying lattice in units of the lattice constant, preserving the original notation.
We emphasize that in distinction to \cite{sachdev 02}, \cite{sachdev QPTs}, where Hamiltonian (\ref{ham1}) was derived, $\Delta_i$ are constants independent of $\mu$ and $J$. We note that Hamiltonian (\ref{ham1}) is not gauge invariant unless $J{=}0$. It is also not hole-particle invariant unless $\mu{=}0$ and $\Delta_i$, $i{=}1,\ldots,d$, are real. The latter condition can be assumed to hold without any loss of generality, since 
Hamiltonian (\ref{ham1}) with any complex $\Delta$ is unitarily equivalent to that with $\Delta$ replaced by $|\Delta|$.

Imposing, independently in each direction ${\bs e}_i$, $i{=}1,\ldots,d$, periodic or antiperiodic boundary conditions,
Hamiltonian (\ref{ham1}) can be simplified by passing from the site-localized to the plane-wave basis labeled by suitable wave vectors (quasimomenta) ${\bs k}{=}(k_1,\ldots,k_d)$,
\begin{equation}
H = \sum_{{\bs k},\sigma}\varepsilon_{\bs k} c^{\dagger}_{{\bs k},\sigma} c_{{\bs k},\sigma}
- J \sum_{{\bs k},i}   \cos{k_i} \left( \Delta_i   c^{\dagger}_{{\bs k},\uparrow}c^{\dagger}_{-{\bs k},\downarrow} + \mathrm{h.c.} \right),
\label{ham2}
\end{equation}
where $\varepsilon_{\bs k}$ stands for the dispersion relation of the hopping term,
\begin{equation}
\varepsilon_{\bs k} = \sum_i \cos k_i - \mu,
\label{epsilon}
\end{equation}
with $k_i{=}2\pi[l_i{-}(L_i{-}1)/2]/L_i$ in the case of periodic boundary condition with an odd $L_i$,  and $k_i{=}\pi[2l_i{-}L_i{+}1]/L_i$
in the case of antiperiodic boundary condition with an even $L_i$.
Formally, Hamiltonian (\ref{ham2}) differs from the well-known BCS Hamiltonian of s-wave superconductivity by the presence of  $\cos k_i$ factors in the gauge-symmetry breaking term. Such Hamiltonians can readily be diagonalized by means of the Bogoliubov transformation. The dispersion relation of quasi-particles reads
\begin{equation}
E_{\bs k} + \sum_{\bs k} \left( \varepsilon_{\bs k} - E_{\bs k} \right),
\label{spectrum}
\end{equation}
where $\sum_{\bs k} \left( \varepsilon_{\bs k} - E_{\bs k} \right)$ is the ground-state energy, and $E_{\bs k}$, given by
\begin{equation}
E_{\bs k} =\sqrt{ \varepsilon_{\bs k}^2 + \left|J \sum_i \Delta_i \cos k_i \right|^2},
\label{Ek}
\end{equation}
are the single quasi-particle energies.
For our choice of boundary conditions, as long as our system is finite the excitation energies $E_{\bs k}$ remain strictly positive: $E_{\bs k} > 0$ for all values of ${\bs k}$, and this is assumed to hold in the sequel.

The Hamiltonian (\ref{ham1}) preserves parity; therefore without any loss of generality we can restrict the state-space to the subspace of even number of particles (electrons). In this subspace, the state $|0\rangle_{qp}$ of an unspecified (but even) number of electrons, defined by
\begin{equation}
|0\rangle_{qp}= \prod_{\bs k}(u_{\bs k} + v_{\bs k} c^{\dagger}_{{\bs k},\uparrow}c^{\dagger}_{-{\bs k},\downarrow})|0\rangle,
\label{gs}
\end{equation}
where $|0\rangle$ is the electron vacuum, with $u_{\bs k}$ real and positive,
\begin{equation}
u_{\bs k}= \sqrt{\frac{1}{2}\left(1+ \frac{\varepsilon_{\bs k}}{E_{\bs k}}\right)},
\label{uk}
\end{equation}
and, in general, complex $v_{\bs k}$,
\begin{equation}
|v_{\bs k}|= \sqrt{\frac{1}{2}\left(1- \frac{\varepsilon_{\bs k}}{E_{\bs k}}\right)}, \,\,\,
\arg v_{\bs k} = \arg \left( J \sum_i \Delta_i \cos k_i \right),
\label{vk}
\end{equation}
is the eigenstate of (\ref{ham2}) to the lowest eigenenergy, $\sum_{\bs k} \left(\varepsilon_{\bs k} - E_{\bs k}  \right)$.
As long as $E_{\bs k} > 0$ for all values of ${\bs k}$, the unique ground state $|0\rangle_{qp}$ is the vacuum of elementary excitations (quasi-particles).
However, on passing to the thermodynamic limit, when the system's linear sizes in all directions tend to infinity, the minimum of $E_{\bs k}$ over ${\bs k}$ (the excitation gap in the spectrum  of quasi-particles) may approach zero at special values of the chemical potential $\mu$ and the coupling constant $J$, and then the ground state becomes degenerate. Those special points in the $(\mu,J)$-plane are the quantum-critical points, where the system undergoes  continuous quantum phase transitions.

All the correlation functions of considered systems can be expressed in terms of two basic two-point correlation functions. Since we are interested only in ground-state correlation functions, taking into account the lattice-translation invariance of our system
these two basic two-point correlation functions can be chosen as follows:
\begin{equation}
_{qp}\langle 0|a^{\dagger}_{{\bs 0},\sigma} a_{{\bs r},\sigma}  |0 \rangle_{qp}  \qquad \textrm{and} \qquad
_{qp}\langle 0|a_{{\bs 0},\sigma} a_{{\bs r},-\sigma}  |0 \rangle_{qp},
\label{corr G g}
\end{equation}
with some $\sigma$.
The first correlation function,
$ _{qp}\langle 0|a^{\dagger}_{{\bs 0},\sigma} a_{{\bs r},\sigma}  |0 \rangle_{qp}$, is gauge and spin-flip invariant;
for ${\bs r}\neq 0$ it represents offdiagonal matrix elements of the ground-state one-body reduced density operator, and amounts to
\begin{equation}
 _{qp}\langle 0|a^{\dagger}_{{\bs 0},\sigma} a_{{\bs r},\sigma}  |0 \rangle_{qp} =
-\frac{1}{2L^d} \sum_{\bs k} \frac{\varepsilon_{\bs k}}{E_{\bs k}} \exp i{\bs k}{\bs r},
\label{corr_G 1}
\end{equation}
which, upon using the invariance of  $\varepsilon_{\bs k}$ and $E_{\bs k}$  with respect to reflections of ${\bs k}$ in coordinate axes, in the thermodynamic limit becomes
\begin{equation}
\lim_{L \to \infty} { _{qp}\langle } 0|a^{\dagger}_{{\bs 0},\sigma} a_{{\bs r},\sigma}  |0 \rangle_{qp}
\equiv G({\bs r}) =
- \frac{1}{2 \pi^d} \int_{0 \leq k_j \leq \pi} d{\bs k} \frac{\varepsilon_{\bs k} }{E_{\bs k}}
\prod_{j=1}^d \cos k_j r_j \,.
\label{corr_G 2}
\end{equation}
Choosing the spin projection $\sigma=\uparrow$, the second correlation function, measuring the degree of gauge-symmetry breaking, amounts to
\begin{equation}
_{qp}\langle 0|a_{{\bs 0},\uparrow} a_{{\bs r},\downarrow}  |0 \rangle_{qp} =
-\frac{1}{2L^d} \sum_{\bs k} \frac{J\sum_i \Delta_i \cos k_i}{E_{\bs k}} \exp i{\bs k}{\bs r},
\label{corr h 1}
\end{equation}
which, by the above arguments, in the thermodynamic limit becomes
\begin{equation}
\lim_{L \to \infty} { _{qp}\langle } 0|a_{{\bs 0},\uparrow} a_{{\bs r},\downarrow}  |0 \rangle_{qp}
\equiv h({\bs r}) =
- \frac{1}{2 \pi^d} \int_{0 \leq k_j \leq \pi} d{\bs k} \frac{J\sum_i \Delta_i \cos k_i}{E_{\bs k}}
\prod_{j=1}^d \cos k_j r_j.
\label{corr_h 2}
\end{equation}

Both the above defined two-point correlation functions are used to define the order parameters in the ground-state phase diagrams presented in the sections that follow. However, analytic results will be given only for the gauge-invariant correlation function $G({\bs r})$, defined in
(\ref{corr_G 2}).
For a fixed lattice direction and the parameters $\Delta_i$, $G({\bs r})$ depends on three parameters: $|{\bs r}|$ - the distance between the two points of the correlation function, the chemical potential $\mu$ and the coupling constant $J$. The ground-state phase diagrams are presented in the $(\mu,J)$-plane, in particular the quantum-critical points of the considered models are uniquely defined by pairs $(\mu,J)$.

In the sequel, we will discuss three kinds of asymptotic behavior of correlation functions, as their variables enter a region specified by the conditions imposed on the variables.
First, the large-distance asymptotic behavior, that is, for a fixed point $(\mu,J)$, distance $|{\bs r}|$ tends to infinity.
Second, the critical-asymptotic behavior, that is, for an arbitrary fixed distance $|{\bs r}|$, $(\mu,J)$-points approach a quantum-critical point along some path in the $(\mu,J)$-plane. Two kinds of paths will be considered: $\mu$-paths that are parallel to the $\mu$-axis and $J$-paths that are parallel to the $J$-axis.
Third, the doubly-asymptotic behavior, that is, for a fixed but sufficiently large $|{\bs r}|$, $(\mu,J)$-points approach a quantum-critical point along a $\mu$-path or a $J$-path.

Anticipating the results presented in the sections that follow, in the one-dimensional model and the two-dimensional symmetric model the critical points are located in a symmetric interval at the $\mu$-axis and at the $J$-axis in the parameter space of $(\mu,J)$-points. In all the analytic asymptotic formulae to be derived, a vicinity of the multicritical point $(0,0)$ is excluded. In particular, for $\mu \to 0$ the $J$-coordinates of  $\mu$-paths have to be away from zero; analogous condition applies to $J$-paths.

In gapped phases, a decay with increasing $|{\bs r}|$ of a two-point correlation function  is dominated by an exponential factor,
$\exp(-|{\bs r}|/\xi)$, which defines the correlation length $\xi$. If additionally $(\mu,J)$-points approach a quantum-critical point, i.e. the distance $\delta$ between them tends to zero (that is in a doubly-asymptotic region), then $\xi$ diverges as $\delta^{-\nu}$, which in turn defines a universal critical index $\nu$ associated with a particular quantum-critical point. Below we demonstrate, providing explicit formulae, that in two-dimensional systems the decay of two-point correlation functions, hence $\xi$ and  $\nu$, depends on lattice directions.

In what follows we shall focus on the gauge-invariant correlation function $G({\bs r})$, defined in (\ref{corr_G 2}). For a reader's convenience, in section \ref{1d} ($d{=}1$-system) and in four subsections of section \ref{2d} (four cases of
$d{=}2$-system), we report the obtained results in the same order. First, we present the large-distance asymptotic behavior of $G({\bs r})$, with explicit expressions for $\xi$. Then, we give the doubly-asymptotic behavior, with the values of $\nu$ shown in the phase diagram, for each kind of quantum-critical points. After that, we derive specific scaling laws, with respect to $\mu$ and $J$, satisfied by $G({\bs r})$ in doubly-asymptotic regions. Finally, we provide numerical arguments that in critical regions, that is for $(\mu,J)$-points sufficiently close to a quantum-critical point, these scaling laws hold for any distances (not only for sufficiently large ones).

\section{\label{1d} The one-dimensional case}

As mentioned in Introduction, this section is included for reasons of completeness.
All the expressions of previous section can be adapted to the one-dimensional case by setting
$\Delta_i {=} k_i{=}r_i=0$ for $i>1$, $\Delta_1 {\equiv} \Delta$, $k_1 {\equiv} k$ and $r_1 {\equiv}  r$.
Moreover, in all the formulae and figures of this section we make the identification $J|\Delta| {\equiv} J$.

We can distinguish four ground-state phases labeled by two order parameters, ${\cal{O}}_1$ and ${\cal{O}}_2$.
The order parameter ${\cal{O}}_1$ is given by
\begin{equation}
{\cal{O}}_1 = G(0) - \frac{1}{2}.
\label{O1}
\end{equation}
When $\Delta$ is real, the system is hole-particle invariant at $\mu=0$, and then ${\cal{O}}_1$ is a deviation of the number density with given spin projection $\sigma$ from its value at the hole-particle symmetry line $\mu{=}0$; it is an odd function of $\mu$. Since $G(r)$ depends only on $|\Delta|$ the latter property of ${\cal{O}}_1$ holds as well if $\Delta$ is complex. The order parameter ${\cal{O}}_2$, measuring the degree of gauge symmetry breaking, is defined as
\begin{equation}
{\cal{O}}_2 = - \Delta^{*} h(1).
\label{O2}
\end{equation}

\begin{figure}
\begin{center}
\includegraphics[width=10cm,clip=on]{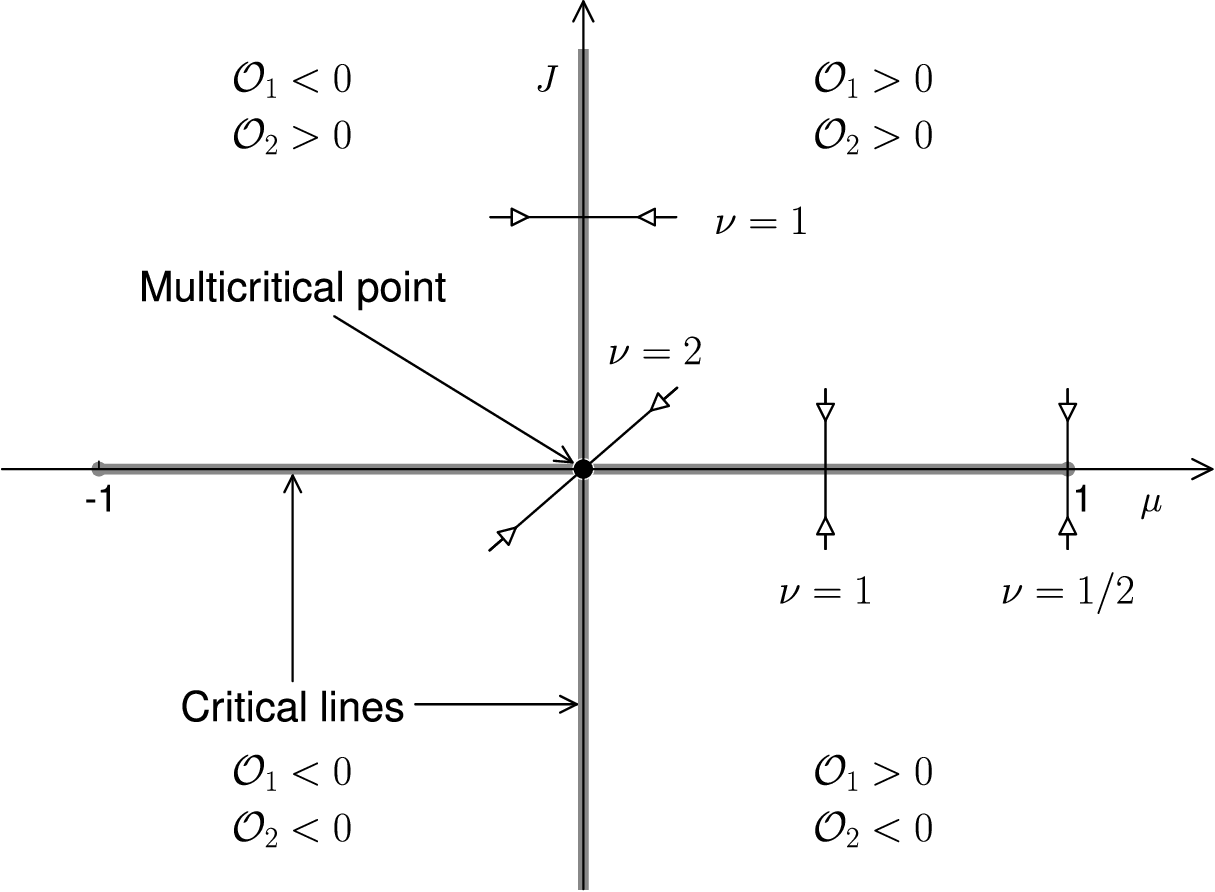}
\caption{\label{diagram1d} Phase diagram of the one-dimensional system in the $(\mu,J)$ plane. The set of quantum critical points consists of the $J$--axis and the closed interval $[-1,1]$ of $\mu$--axis -- thick lines. Those lines constitute also phase boundaries of the four phases, labeled by the order parameters ${\cal{O}}_1$ and ${\cal{O}}_2$. Double arrows indicate the types of neighborhoods of critical points, in which the asymptotic behaviors of $G(r)$ are studied, except neighborhoods of the multicritical point.  By the arrows, the values of the universal critical indices $\nu$ in those neighborhoods are given.}
\end{center}
\end{figure}

The quantum critical points of the one-dimensional system are located at the $J$--axis  and in the closed interval $[-1,1]$ of the $\mu$--axis. There are two critical end points $(\pm 1,0)$ and a multicritical point $(0,0)$.
The ground-state phase diagram of the one-dimensional system is shown in Fig.~\ref{diagram1d}.

\begin{figure}
\begin{center}
\includegraphics[width=15cm,clip=on]{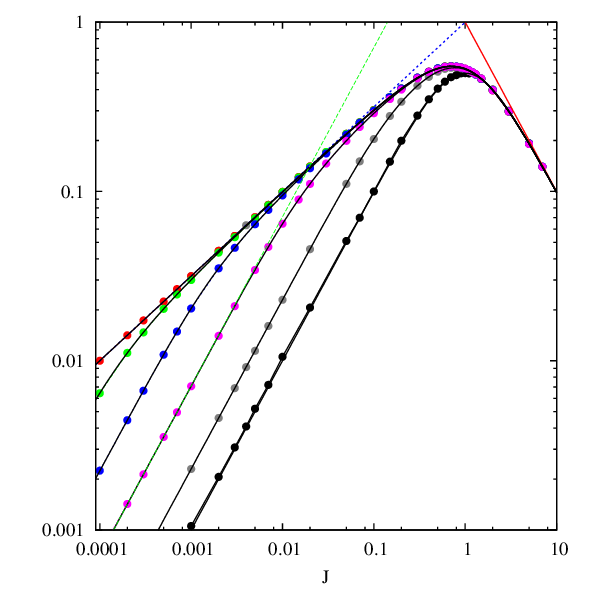}
\caption{\label{1d xi} (Color online) $d{=}1$. Plots of $(\mu \xi)^{-1}$ versus $J$ in doubly logarithmic scale, for different values of $\mu$.  Color balls represent the values calculated numerically from the large-distance behavior of $G(r)$; from top to bottom: $\mu{=}1.0000$ -- red, $\mu{=}0.9999$ -- green, $\mu{=}0.9990$ -- blue,
$\mu{=}0.9900$ -- magenta, $\mu{=}0.9000$ -- gray and $\mu{=}0.2000$ -- black.
Black-continuous lines  are obtained from formula (\ref{xi theta phi}). The dashed lines are the asymptotic lines as $J \to 0$.
For $\mu=1$, it is the blue-dashed line, which is the plot of $\sqrt{J}$. For $0 {<}\mu {<}1$, the asymptotic lines are  the plots of $J/\sqrt{1-\mu^2}$;  green-dashed line is a representative plot for $\mu=0.9900$.
The red-continuous line, which is the plot of $J^{-1}$, is the asymptotic line as $J \to \infty$, for any
$0 {<}\mu {\leq} 1$.}
\end{center}
\end{figure}

We start with presenting a summary of analytic results concerning the large-distance asymptotic behavior of $G(r)$.
It is convenient to introduce an auxiliary function ${\tilde{G}}(r)$ :
\begin{eqnarray}
{\tilde{G}}(r) = - \frac{1}{2\pi}\int_{0}^{\pi}dk\frac{\cos(rk)}{\sqrt{(\cos k - \mu)^2+J^2\cos^2k}},
\label{corr G 1 tilde}
\end{eqnarray}
in terms of which
\begin{equation}
G(r) = \frac{1}{2} \left[ {\tilde{G}}(r+1) + {\tilde{G}}(r-1) \right]
- \mu {\tilde{G}}(r).
\label{corr G 3}
\end{equation}
In the stripe $|\mu| \leq 1$ of the $(\mu,J)$--plane, but excluding the $\mu{=}0$ and $J{=}0$ lines of quantum critical points, the large-distance asymptotic behavior of ${\tilde{G}}(r)$, derived by heuristic arguments, reads
\begin{eqnarray}
{\tilde{G}}(r) \approx - \sqrt{\frac{1+J^2}{2\pi |\mu J| c^{1/4}}} \frac{\exp(-r/\xi)}{\sqrt{r}}
\cos(\theta r-\phi),
\label{corr G asympt}
\end{eqnarray}
where
\begin{equation}
\frac{1}{\xi}=\ln\left(a_{+} + \sqrt{a_{+}^2-1}\right), \quad
\theta=\arccos (a_{-}), \quad
\phi = \frac{1}{4}\arctan\frac{2\mu|\mu J|}{(1+J^2)^2-\mu^2(1-J^2)},
\label{xi theta phi}
\end{equation}
with
\begin{equation}
a_{\pm} = \frac{1}{2(1+J^2)}\left[\sqrt{(\mu+1+J^2)^2+\mu^2J^2} \pm  \sqrt{(\mu-1-J^2)^2+\mu^2J^2}\right],
\label{a}
\end{equation}
and
\begin{equation}
c = \left[(1+J^2)^2-\mu^2(1-J^2)\right]^2+4\mu^4J^2.
\label{c}
\end{equation}
We note that the correlation length, given by formulae (\ref{xi theta phi}), (\ref{a}) and (\ref{c}), is in excellent agreement with the correlation length determined numerically, which is demonstrated in Fig.~\ref{1d xi}.

As might be expected, the doubly asymptotic formulae (i.e. holding in a doubly asymptotic region of sufficiently large distances and $(\mu,J)$-points sufficiently close to a quantum critical point) are considerably simpler than the large-distance asymptotic formula (\ref{corr G asympt}).
Specifically, in doubly asymptotic regions, where $(\mu,J)$-points approach along a $\mu$-path a point belonging to any one of the two half lines of quantum critical points, $\mu{=}0$ and $J{\neq} 0$, formula (\ref{corr G asympt}) gives
\begin{equation}
|{\tilde{G}}(r)| \approx
\sqrt{\frac{1}{2\pi |\mu J|}}\frac{\exp(-r/\xi)}{\sqrt{r}} \left| \cos r\left( \frac{\pi}{2}-\frac{1}{|J|\xi} \right)\right|,
\textrm{with} \qquad \frac{1}{\xi} \approx \frac{|\mu J|}{1+J^2},
\label{asympt mu=0}
\end{equation}
then if $(\mu,J)$-points approach along the $J$-path one of the two end critical points, $|\mu|{=}1$ and $J{=}0$,
\begin{equation}
|{\tilde{G}}(r)| \approx
\left( \frac{1}{\pi^2 |2J|^3} \right)^{1/4}\frac{\exp(-r/\xi)}{\sqrt{r}} \left| \cos\left(\frac{r}{\xi} -\frac{\pi}{8} \right)\right|,
\textrm{with} \qquad \frac{1}{\xi} \approx \sqrt{|J|},
\label{asympt mu=1}
\end{equation}
and finally if $(\mu,J)$-points approach along a $J$-path any point of the two line segments of quantum critical points, $J{=}0$ and $0 {<} |\mu| {<} 1$,
\begin{eqnarray}
|{\tilde{G}}(r)| \approx
\sqrt{\frac{1}{2\pi |\mu J|(1-\mu^2)^{1/2}}} \frac{\exp\left(-r/\xi\right)}{\sqrt{r}} \times \nonumber  \\
\left|\cos\left[\left(\frac{\pi}{2}-\arcsin|\mu| + \frac{|\mu| (2-\mu^2)J^2}{2(1-\mu^2)^{3/2}} \right)r
-\frac{1}{2|J|\xi^2}\right]\right|, \quad
\textrm{with} \quad \frac{1}{\xi} \approx \frac{|\mu J|}{\sqrt{1-\mu^2}}.
\label{asympt J=0}
\end{eqnarray}
From the expressions for the correlation lengths in neighborhoods of quantum critical points, given in (\ref{asympt mu=0}), (\ref{asympt mu=1}) and (\ref{asympt J=0}), one readily obtains the values of indices $\nu$ displayed in the phase diagram, Fig.~\ref{diagram1d}. We note also, that irrespectively of the value of $\mu$ (but $\mu$ separated from $0$), $\xi^{-1}$ tends asymptotically to $|\mu/J|$ as $|J|{\to}\infty$, see
Fig.~\ref{1d xi}.

Apparently, in each one of the above doubly asymptotic formulae there are three factors: a distance-independent positive coefficient ${\cal{C}}$, a damping factor ${\cal{D}}$, determining decay of correlations with distance (a product of an exponential and power factors), and an oscillating factor ${\cal{O}}$, so that $|{\tilde{G}}(r)| \approx {\cal{C}}{\cal{D}}{\cal{O}}$.
We note that in doubly asymptotic regions such factorizations hold as well for the whole correlation function $G(r)$.
An inspection of doubly asymptotic formulae (\ref{asympt mu=0}), (\ref{asympt mu=1}) and (\ref{asympt J=0}) for the auxiliary function ${\tilde{G}}(r)$ reveals interesting scaling properties, with respect to distance and one of the parameters, $\mu$ or $J$, of the correlation function $G(r)$ or of some of its factors. Specifically, in the corresponding doubly asymptotic regions, formula (\ref{asympt mu=0}) implies
\begin{equation}
|G(r)| \approx |\mu| \, g_{J}(|\mu| r),
\label{scal mu r}
\end{equation}
then from (\ref{asympt mu=1}) we derive
\begin{equation}
|G(r)| \approx \sqrt{|J|} \, g_{|\mu|=1}(\sqrt{|J|} r),
\label{scal J r mu=1}
\end{equation}
and finally, formula (\ref{asympt J=0}) gives
\begin{equation}
{\cal{C}}{\cal{D}} \approx |J| \, g_{\mu}(|J| r),
\label{scal J r}
\end{equation}
where $g_J$, $g_{\mu}$ and $g_{|\mu|=1}$ stand for some functions.
From scaling formulae (\ref{scal mu r}), (\ref{scal J r mu=1}) and (\ref{scal J r}) one can infer the values of critical indices $\nu$ in the corresponding critical regions: $1$, $1/2$ and $1$, respectively.

\begin{figure}
\centering
\begin{subfigure}[b]{0.48\textwidth}
\centering
\includegraphics[width=8cm,clip=on]{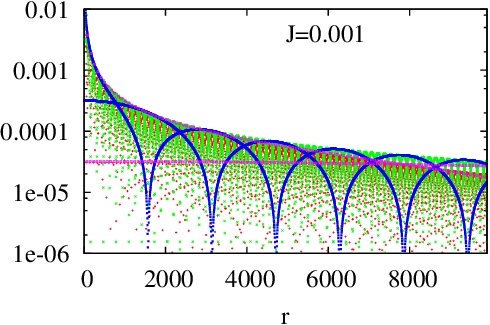}
\end{subfigure}
~
\begin{subfigure}[b]{0.48\textwidth}
\centering
\includegraphics[width=8cm,clip=on]{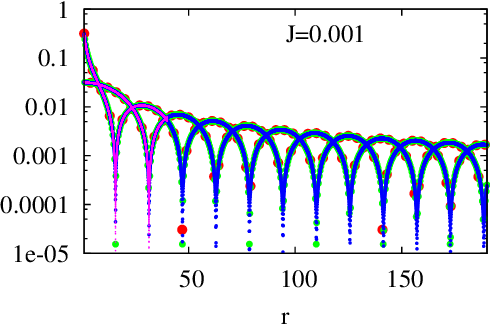}
\end{subfigure}
\caption{\label{cf_J0001}\label{cf_J0001_sc}(Color online) $d{=}1$. Left panel, plots of $|G(r)|$ in logarithmic scale
for $J{=}0.001$ and four values of $\mu$: $\mu{=}0.1$ -- red line, $\mu{=}0.01$ -- green line, $\mu{=}0.001$ -- blue line, $\mu{=}0.0001$ -- magenta line. Right panel, the plot for $\mu{=}0.01$ is repeated, while the remaining ones are scaled according to formula (\ref{scal' mu r}). Clearly, all the plots merge into one plot in the whole range of distances.}
\end{figure}

\begin{figure}
\centering
\begin{subfigure}[b]{0.48\textwidth}
\centering
\includegraphics[width=8cm,clip=on]{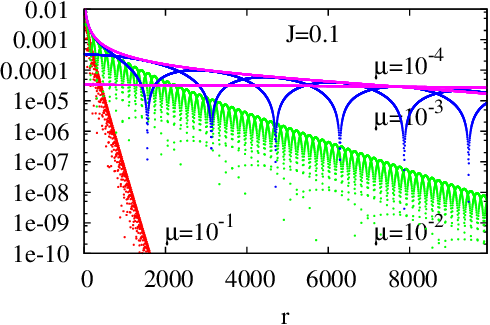}
\end{subfigure}
~
\begin{subfigure}[b]{0.48\textwidth}
\centering
\includegraphics[width=8cm,clip=on]{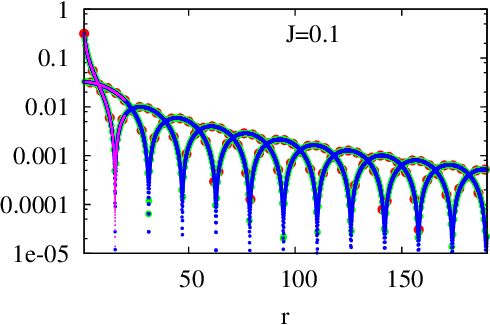}
\end{subfigure}
\caption{\label{cf_J01}\label{cf_J01_sc}(Color online) $d{=}1$. Left panel, plots of $|G(r)|$ in logarithmic scale
for $J{=}0.1$ and four values of $\mu$: $\mu{=}0.1$ -- red line, $\mu{=}0.01$ -- green line, $\mu{=}0.001$ -- blue line, $\mu{=}0.0001$ -- magenta line. Right panel, the plot for $\mu{=}0.01$ is repeated, while the remaining ones are scaled according to formula (\ref{scal' mu r}). Clearly, all the plots merge into one plot in the whole range of distances.}
\end{figure}

\begin{figure}
\centering
\begin{subfigure}[b]{0.48\textwidth}
\centering
\includegraphics[width=8cm,clip=on]{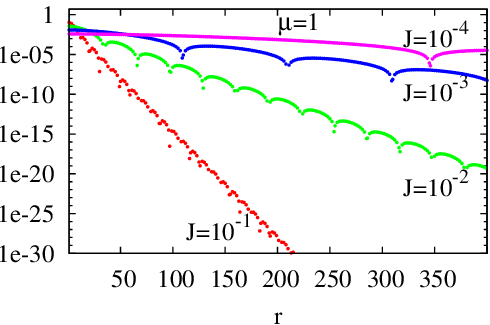}
\end{subfigure}
~
\begin{subfigure}[b]{0.48\textwidth}
\centering
\includegraphics[width=8cm,clip=on]{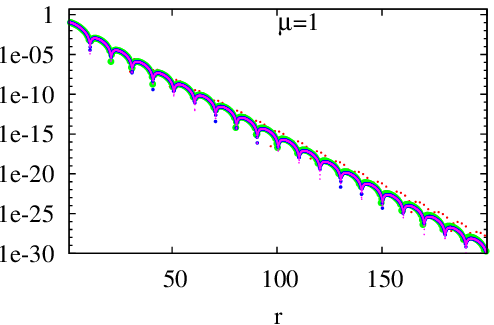}
\end{subfigure}
\caption{\label{cf_mu1}\label{cf_mu1_sc}(Color online) $d{=}1$. Left panel, plots of $|G(r)|$ in logarithmic scale
for $\mu{=}1$ and four values of $J$: $J{=}0.1$ -- red line, $J{=}0.01$ -- green line, $J{=}0.001$ -- blue line, $J{=}0.0001$ -- magenta line. Right panel, the plot for $J{=}0.01$ is repeated, while the remaining ones are scaled according to formula (\ref{scal J r mu=1}).  Clearly, all the plots merge into one plot in the whole range of distances.
It is well seen that scaling properties hold only sufficiently close to the critical point; $J{=}0.1$ is too large and the corresponding plot does not coincide with the other ones.}
\end{figure}

\begin{figure}
\centering
\begin{subfigure}[b]{0.48\textwidth}
\centering
\includegraphics[width=8cm,clip=on]{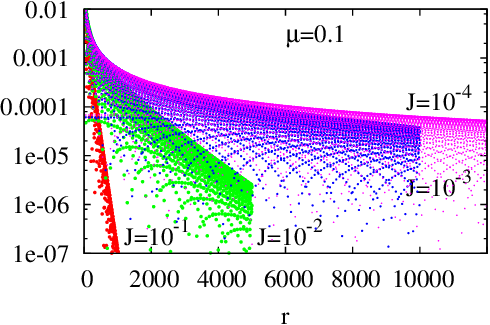}
\end{subfigure}
~
\begin{subfigure}[b]{0.48\textwidth}
\centering
\includegraphics[width=8cm,clip=on]{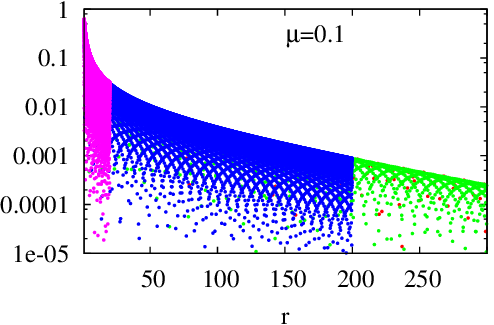}
\end{subfigure}
\caption{\label{cf_mu01_}\label{cf_mu01_sc}(Color online) $d{=}1$. Left panel, plots of $|G(r)|$ in logarithmic scale
for $\mu{=}0.1$ and four values of $J$: $J{=}0.1$ -- red line, $J{=}0.01$ -- green line, $J{=}0.001$ -- blue line, $J{=}0.0001$ -- magenta line. Right panel, the plot for $J{=}0.01$ is repeated, while the remaining ones are scaled according to formula (\ref{scal J r}).
Apparently,  the whole function $G(r)$ does not scale according to (\ref{scal J r}). Clearly, only the envelops of the plots of $|G(r)|$, describing decay of correlations with distance, merge into one plot in the whole range of distances.}
\end{figure}

Numerical calculations of $G(r)$ reveal a remarkable fact. Namely, the scaling laws (\ref{scal mu r}), (\ref{scal J r mu=1}) and (\ref{scal J r}), derived only in doubly asymptotic regions, hold in the whole range of distances; only a proximity of $(\mu,J)$-point to a quantum critical one is required.
To demonstrate this, consider for definiteness the scaling of $|G(r)|$ with respect to $\mu$ and $r$. Let us write scaling formula (\ref{scal mu r}) as $|G_{\mu}(r)| \approx |\mu| g_{J}(|\mu| r)$.
For two values of the chemical potential, $\mu$ and $\lambda \mu$, $\lambda$ -- a positive number, we readily find the relation
\begin{equation}
\lambda^{-1} |G_{\lambda \mu}(r)| \approx |G_{\mu}(\lambda r)|.
\label{scal' mu r}
\end{equation}
In Figs.~\ref{cf_J0001} and \ref{cf_J01}, we demonstrate that relation (\ref{scal' mu r}) holds surprisingly well even for relatively small distances.
In an analogous way we verified the scaling formula (\ref{scal J r mu=1}), see Fig.~\ref{cf_mu1},
and (\ref{scal J r}), see Fig.~\ref{cf_mu01_}.

\section{\label{2d} The two-dimensional case}

All the expressions of section \ref{models} can be adapted to the two-dimensional case by setting $\Delta_i{=}k_i{=}r_i{=}0$ for $i>2$.
But in distinction to the one-dimensional case, due to the freedom in choosing the relation between the parameters $\Delta_1$ and $\Delta_2$, formula (\ref{ham1}) represents a great variety of models. In this paper we limit our considerations to only two cases. Namely, the symmetric case, with the interaction term invariant under rotations by $\pi/2$, where $\Delta_1 {=} \Delta_2 {=} \Delta$, and the antisymmetric case, with the interaction term that changes sign under a rotation by $\pi/2$, where $\Delta_1 {=} - \Delta_2 {=} \Delta$. We note that in both cases the correlation functions of our systems are invariant not only with respect to lattice translations but also with respect to rotations by $\pi/2$.

As compared to the one-dimensional case, a novel feature of two-dimensional models is that the two-point correlation functions $G({\bs r})$ depend not only on the distance $|{\bs r}|$ but also on the direction of ${\bs r}$. Expressing ${\bs r}$ by its Cartesian coordinates, ${\bs r}{=}(r_1,r_2)$, we can parameterize directions by the ratio $r_1/r_2 \equiv n$.
Then, for a given critical point, we can expect $n$-dependent doubly-asymptotic behaviors of correlations. Unfortunately, the analytic asymptotic formulae for $G({\bs r})$, which we have been able to obtain, apply only to points ${\bs r}$ such that $n \geq n_0 >1$ or, by symmetry, $n \leq n^{-1}_0 <1$, that is for offdiagonal directions which form a sufficiently small angle with the axial directions. Therefore, the asymptotic formulae in the diagonal direction are derived separately.
These formulae define $n$-dependent correlation lengths $\xi^{(\pm)}_{\text{offdiag}}$ in offdiagonal directions satisfying the conditions specified above and the correlations length $\xi^{(\pm)}_{\text{diag}}$ in the diagonal direction, where the superscript plus refers to the symmetric model and minus -- to the antisymmetric one.
Interestingly, our analytical and numerical results show that, for each critical point, there are only two kinds of universal critical indices $\nu$: $\nu_{\text{offdiag}}$ for all offdiagonal directions and $\nu_{\text{diag}}$ for the diagonal direction.

Similarly to the one-dimensional case, in the formulae and figures of this section we make the identification
$J|\Delta| {\equiv} J$,

\subsection{The symmetric model}

We can distinguish four ground-state phases labeled by two order parameters, ${\cal{O}}_1$ and ${\cal{O}}_2$.
The parameter ${\cal{O}}_1$ is defined as in the one-dimensional case, formula (\ref{O1}), while the new definition of ${\cal{O}}_2$ is
\begin{equation}
{\cal{O}}_2 = - \Delta^{*}h(1,0).
\label{O2}
\end{equation}
The comments of previous section related to  ${\cal{O}}_1$ and ${\cal{O}}_2$ remain valid.

\begin{figure}
\begin{center}
\includegraphics[width=10cm,clip=on]{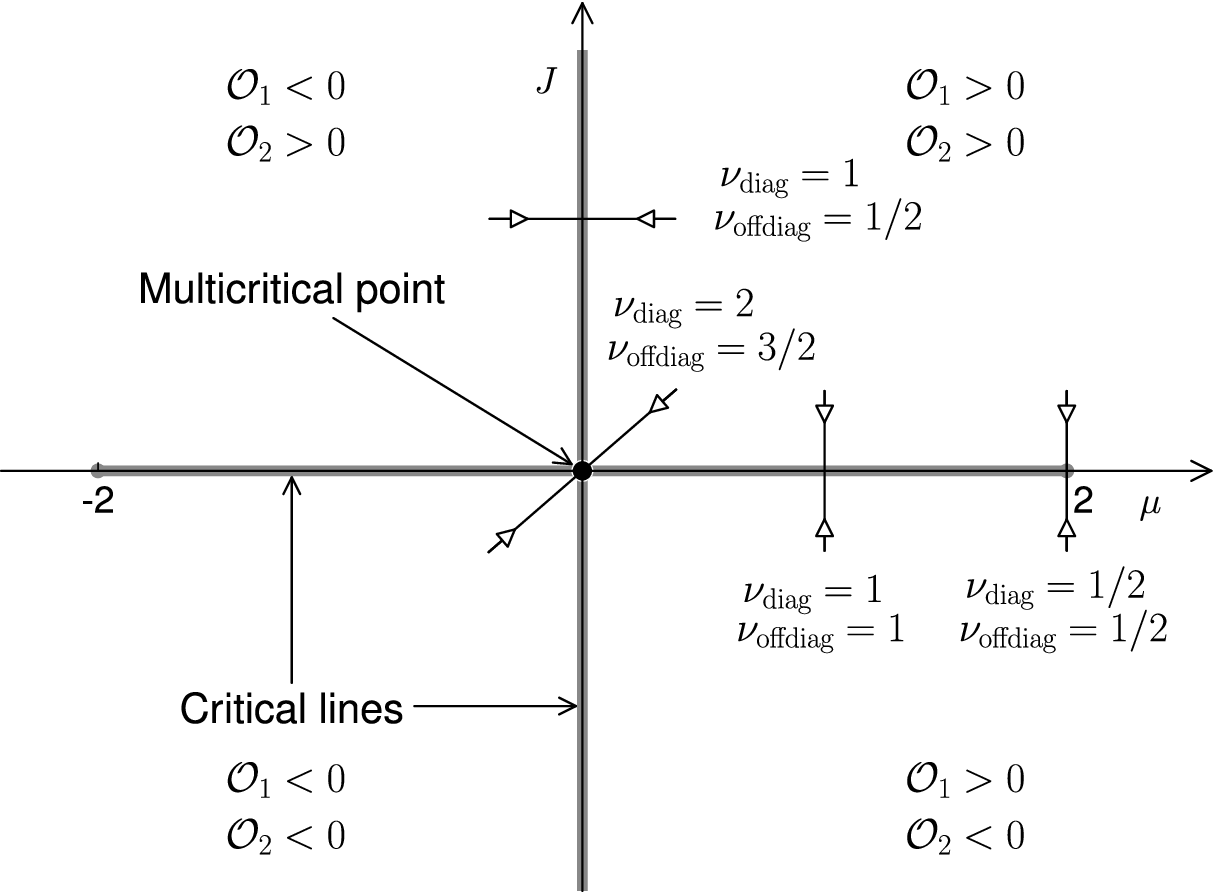}
\caption{\label{diagram2d+} Phase diagram of the symmetric two-dimensional system in the $(\mu,J)$--plane. The set of quantum-critical points consists of the $J$--axis and the closed interval $[-2,2]$ of $\mu$--axis -- thick lines. Those lines constitute also phase boundaries of the four phases, labeled by the order parameters ${\cal{O}}_1$ and ${\cal{O}}_2$. Double arrows indicate the types of neighborhoods of critical points, in which the asymptotic behaviors of $G({\bs r})$ are studied, except neighborhoods of the multicritical point. The universal critical indices $\nu$ in those neighborhoods, whose values are given by the arrows, depend in general on lattice direction, whether it is diagonal ($\nu_{\text{diag}}$) or offdiagonal ($\nu_{\text{offdiag}}$).}
\end{center}
\end{figure}

The quantum-critical points of the symmetric two-dimensional system are located at the $J$--axis  and in the closed interval $[-2,2]$ of the $\mu$--axis. There are two critical end points $(\pm 2,0)$ and a multicritical point $(0,0)$.
The ground-state phase diagram of the symmetric two-dimensional system is shown in Fig.~\ref{diagram2d+}.

{{\bf (i) the diagonal direction}}: $r_1{=}r_2{=}r^{\prime}$ \\

\begin{figure}
\begin{center}
\includegraphics[width=15cm,clip=on]{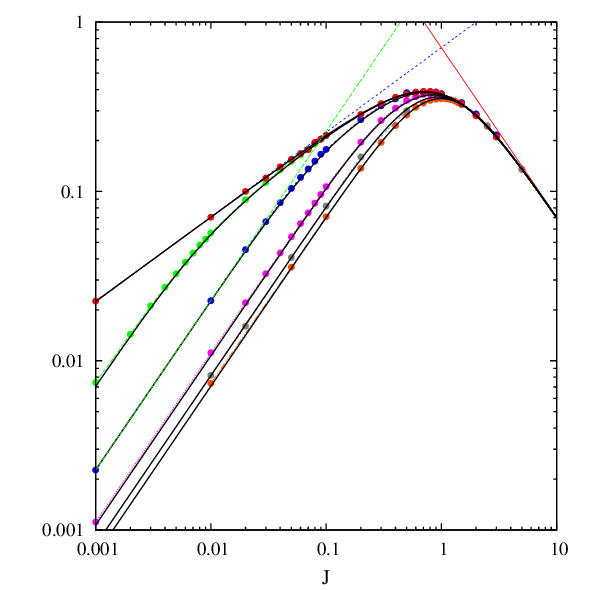}
\caption{\label{diag_2d+ xi} (Color online) $d{=}2$, the symmetric model and the diagonal direction. Plots of $(\mu \xi_{\text{diag}}^{(+)})^{-1}$ versus $J$ in doubly logarithmic scale, in diagonal direction, for different values of $\mu$. Color balls represent the values calculated numerically from the large-distance behavior of ${G}(r^{\prime},r^{\prime})$;
from top to bottom: $\mu{=}2.00$ -- red, $\mu{=}1.99$ -- green, $\mu{=}1.90$ -- blue, $\mu{=}1.50$ -- magenta,
$\mu{=}1.00$ -- gray and $\mu{=}0.10$ -- orange.
Black-continuous lines  are obtained from formulae (\ref{2d+_diag_xi theta phi}) and (\ref{2d+_diag_a}).
The dashed lines are the asymptotic lines as $J \to 0$. For $\mu{=}2$, it is the blue-dashed line, which is the plot of $\sqrt{J/2}$. For $0{<}\mu{<}2$, the asymptotic lines are the plots of $\sqrt{2/(4-\mu^2)}J$;
green-dashed line is a representative plot for $\mu{=}1.90$.
The red-continuous line, which is the plot of $(\sqrt{2}J)^{-1}$, is the asymptotic line as $J \to \infty$, for any
$0 {<}\mu {\leq} 2$.}
\end{center}
\end{figure}

The correlation function $G(r^{\prime},r^{\prime})$ is given by the general formula (\ref{corr_G 2}),
\begin{eqnarray}
{G}(r^{\prime},r^{\prime})= - \frac{1}{2\pi^2}\int_{0 \leq k_1,k_2 \leq \pi}dk_1 dk_2
\cos(r^{\prime}k_1)\cos(r^{\prime}k_2)
\frac{\varepsilon_{(k_1,k_2)}}{E_{(k_1,k_2)}},
\label{G_diag_2d+}
\end{eqnarray}
where
\begin{equation}
\varepsilon_{(k_1,k_2)} = \cos k_1 + \cos k_2 - \mu, \,\,\,
E_{(k_1,k_2)} = \sqrt{(\cos k_1 + \cos k_2 -\mu)^2 + J^2(\cos k_1 + \cos k_2)^2}.
\label{eps E 2d+}
\end{equation}
Naturally, determining its large-distance asymptotic behavior is a harder task, then in the one-dimensional case, considered in section \ref{1d}.
In the stripe $|\mu| \leq 2$ of the $(\mu,J)$--plane, but excluding the $\mu{=}0$ and $J{=}0$ lines, the large-distance asymptotic behavior of $G(r^{\prime},r^{\prime})$ is given by
\begin{eqnarray}
{G}(r^{\prime},r^{\prime}) \approx - \textrm{sgn} (\mu)\frac{1}{2\pi} \left( \frac{J^2}{1+J^2} \right)^{1/4}
\frac{\exp(-r^{\prime}/\xi^{(+)})}{r^{\prime}} \cos(\theta r^{\prime}+\phi),
\label{G_diag_asympt_2d+}
\end{eqnarray}
where
\begin{equation}
\frac{1}{\xi^{(+)}}=2\ln\left(a_{+} + \sqrt{a_{+}^2-1}\right), \qquad
\theta=2\arccos (a_{-}), \qquad
\phi =\frac{\pi}{4}-\frac{1}{2}\arctan |J|,
\label{2d+_diag_xi theta phi}
\end{equation}
\begin{equation}
a_{\pm} \equiv \frac{1}{2(1+J^2)}\left[\sqrt{(|\mu|/2+1+J^2)^2+\mu^2J^2/4} \pm \sqrt{(|\mu|/2-1-J^2)^2+\mu^2J^2/4}\right].
\label{2d+_diag_a}
\end{equation}
Note that, since the distance between the points $(0,0)$ and $(r^{\prime},r^{\prime})$ is
$r{=}\sqrt{2}r^{\prime}$, $\xi^{(+)}$ is not the correlation length; the correlation length in the diagonal direction
$\xi_{\text{diag}}^{(+)} {=} \sqrt{2}\xi^{(+)}$.
The formulae (\ref{2d+_diag_xi theta phi}) and (\ref{2d+_diag_a}) for the correlation length $\xi_{\text{diag}}$,  are in excellent agreement with correlation lengths determined numerically, which is demonstrated in Fig.~\ref{diag_2d+ xi}.

The large-distance asymptotic formula (\ref{G_diag_asympt_2d+}) assumes a lot simpler form in doubly asymptotic regions; we give only simplified expressions for $\xi^{(+)}$ and $\theta$.
Specifically, in doubly asymptotic regions, where $(\mu,J)$-points approach along a $\mu$-path a point belonging to any one of the two half lines of quantum critical points, $\mu{=}0$ and $J{\neq} 0$, formulae (\ref{G_diag_asympt_2d+}),
(\ref{2d+_diag_xi theta phi}) and (\ref{2d+_diag_a}) imply
\begin{equation}
\frac{1}{\xi^{(+)}} \approx \frac{|\mu J|}{1+J^2}, \qquad
\theta \approx \pi-\frac{|\mu|}{1+J^2},
\label{diag_asympt 2d+ mu=0}
\end{equation}
then if $(\mu,J)$-points approach along the $J$-path one of the two end critical points, $|\mu|{=}2$ and $J{=}0$,
\begin{equation}
\frac{1}{\xi^{(+)}} \approx 2|J|^{1/2}, \qquad
\theta \approx2|J|^{1/2},
\label{diag_asympt 2d+ mu=2}
\end{equation}
and finally if $(\mu,J)$-points approach along a $J$-path any point of the two line segments of quantum critical points, $J{=}0$ and $0< |\mu| <2$,
\begin{eqnarray}
\frac{1}{\xi^{(+)}} \approx\frac{2|\mu J|}{\sqrt{4-\mu^2}}, \qquad
\theta \approx 2\arccos\frac{|\mu|}{2}.
\label{diag_asympt 2d+ J=0}
\end{eqnarray}
From the expressions for the correlation lengths in neighborhoods of quantum critical points, given in (\ref{diag_asympt 2d+ mu=0}),
(\ref{diag_asympt 2d+ mu=2}) and (\ref{diag_asympt 2d+ J=0}), one readily obtains the values of indices $\nu$ displayed in the phase diagram,
Fig.~\ref{diagram2d+}. We note also that as $|J|{\to}\infty$, $(\xi_{\text{diag}}^{(+)})^{-1}$ tends to $|\mu/\sqrt{2}J|$, irrespectively of $\mu$ separated from zero, clarify Fig.~\ref{diag_2d+ xi}.

\begin{figure}
\centering
\begin{subfigure}[b]{0.48\textwidth}
\centering
\includegraphics[width=8cm,clip=on]{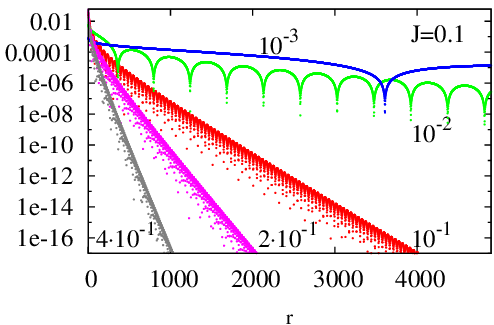}
\end{subfigure}
~
\begin{subfigure}[b]{0.48\textwidth}
\centering
\includegraphics[width=8cm,clip=on]{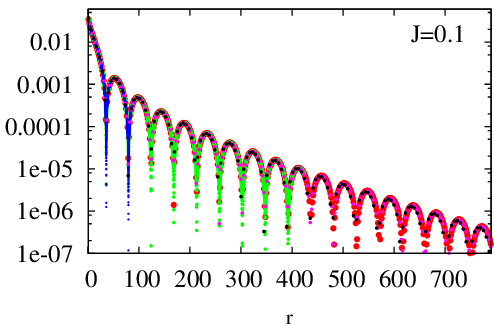}
\end{subfigure}
\caption{\label{cf_J01_ry1rx}\label{cf_J01_ry1rx_sc}(Color online) $d{=}2$, the symmetric model and the diagonal direction. Left panel, plots of $|G(r^{\prime},r^{\prime})|$  versus distance $r$, in logarithmic scale,
for $J{=}0.1$ and five values of $\mu$: $\mu{=}0.1$ -- red line, $\mu{=}0.01$ -- green line, $\mu{=}0.001$ -- blue line, $\mu{=}0.2$ -- magenta line, $\mu{=}0.4$ -- black line. Right panel, the plot for $\mu{=}0.1$ is repeated, while the remaining ones are scaled according to
formula (\ref{scal mu r r}). Clearly, all the plots merge into one plot in the whole range of distances.}
\end{figure}

\begin{figure}
\centering
\begin{subfigure}[b]{0.48\textwidth}
\centering
\includegraphics[width=8cm,clip=on]{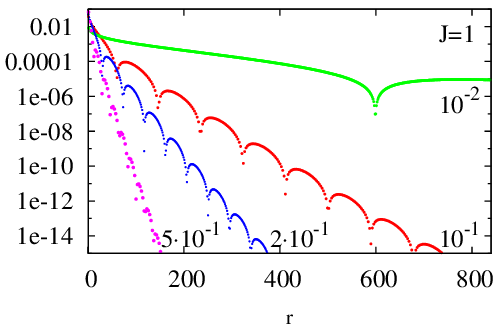}
\end{subfigure}
~
\begin{subfigure}[b]{0.48\textwidth}
\centering
\includegraphics[width=8cm,clip=on]{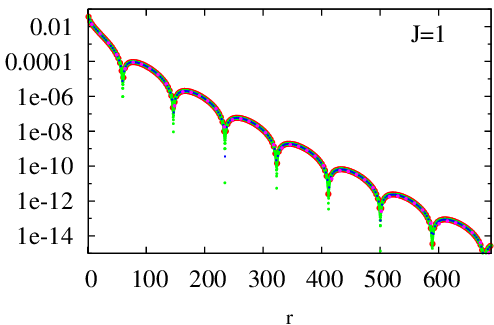}
\end{subfigure}
\caption{\label{cf_J1_ry1rx}\label{cf_J1_ry1rx_sc}(Color online) $d{=}2$, the symmetric model and the diagonal direction. Left panel, plots of $|G(r^{\prime},r^{\prime})|$  versus distance $r$, in logarithmic scale,
for $J{=}1$ and four values of $\mu$: $\mu{=}0.1$ -- red line, $\mu{=}0.01$ -- green line, $\mu{=}0.2$ -- blue line, $\mu{=}0.5$ -- magenta line. Right panel, the plot for $\mu{=}0.1$ is repeated, while the remaining ones are scaled according to formula (\ref{scal mu r r}). Clearly, all the plots merge into one plot in the whole range of distances.}
\end{figure}

\begin{figure}
\centering
\begin{subfigure}[b]{0.48\textwidth}
\centering
\includegraphics[width=8cm,clip=on]{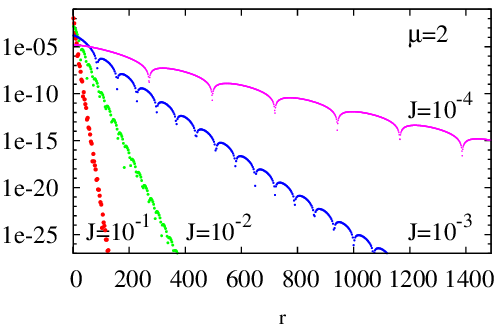}
\end{subfigure}
~
\begin{subfigure}[b]{0.48\textwidth}
\centering
\includegraphics[width=8cm,clip=on]{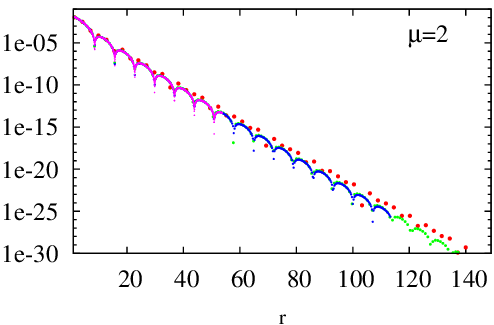}
\end{subfigure}
\caption{\label{cf_mu2_ry1rx}\label{cf_mu2_ry1rx_sc}(Color online) $d{=}2$, the symmetric model and the diagonal direction. Left panel, plots of $|G(r^{\prime},r^{\prime})|$  versus distance $r$, in logarithmic scale,
for $\mu{=}2$ and four values of $J$: $J{=}0.1$ -- red line, $J{=}0.01$ -- green line, $J{=}0.001$ -- blue line, $J{=}0.0001$ -- magenta line. Right panel, the plot for $J{=}0.1$ is repeated, while the remaining ones are scaled according to formula (\ref{scal J r r mu=2}). Clearly, all the  plots merge into one plot in the whole range of distances. It is well seen that scaling properties hold only sufficiently close to the critical point; $J{=}0.1$ is too large and the corresponding plot does not coincide with the other ones.}
\end{figure}

\begin{figure}
\centering
\begin{subfigure}[b]{0.48\textwidth}
\centering
\includegraphics[width=8cm,clip=on]{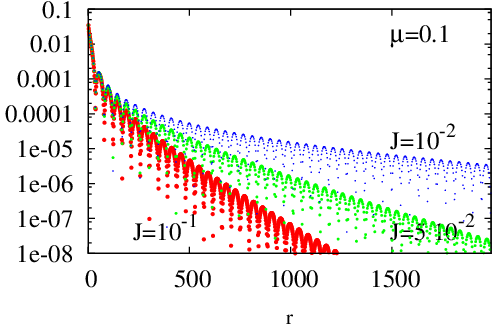}
\end{subfigure}
~
\begin{subfigure}[b]{0.48\textwidth}
\centering
\includegraphics[width=8cm,clip=on]{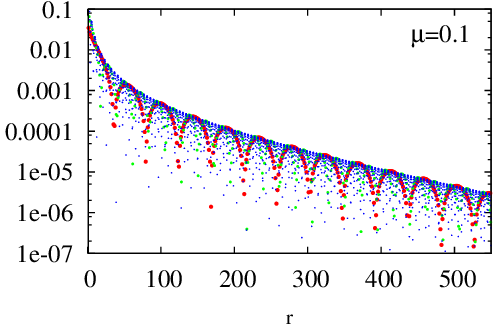}
\end{subfigure}
\caption{\label{cf_mu01_ry1rx}\label{cf_mu01_ry1rx_sc}(Color online) $d{=}2$, the symmetric model and the diagonal direction. Left panel, plots of $|G(r^{\prime},r^{\prime})|$  versus distance $r$, in logarithmic scale,
for $\mu{=}0.1$ and three values of $J$: $J{=}0.1$ -- red line, $J{=}0.05$ -- green line, $J{=}0.01$ -- blue line. Right panel, the plot for $J{=}0.1$ is repeated, while the remaining ones are scaled according to formula (\ref{scal J r r}). Clearly, only the envelops of the plots of $|G(r^{\prime},r^{\prime})|$, describing decay of correlations with distance $r$, merge into one plot in the whole range of distances.}
\end{figure}

\begin{figure}
\centering
\begin{subfigure}[b]{0.48\textwidth}
\centering
\includegraphics[width=8cm,clip=on]{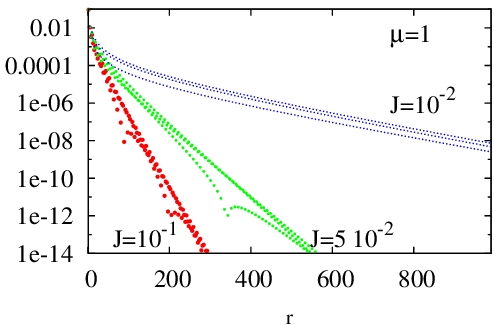}
\end{subfigure}
~
\begin{subfigure}[b]{0.48\textwidth}
\centering
\includegraphics[width=8cm,clip=on]{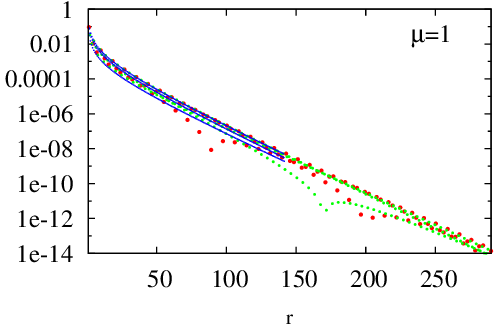}
\end{subfigure}
\caption{\label{cf_mu1_ry1rx}\label{cf_mu1_ry1rx_sc}(Color online) $d{=}2$, the symmetric model and the diagonal direction. Left panel, plots of $|G(r^{\prime},r^{\prime})|$  versus distance $r$, in logarithmic scale,
for $\mu{=}1$ and three values of $J$: $J{=}0.1$ -- red line, $J{=}0.05$ -- green line, $J{=}0.01$ -- blue line. Right panel, the plot for $J{=}0.1$ is repeated, while the remaining ones are scaled according to formula (\ref{scal J r r}). Clearly, only the envelops of the plots of $|G(r^{\prime},r^{\prime})|$, describing decay of correlations with distance $r$, merge into one plot in the whole range of distances. }
\end{figure}

As in the one-dimensional case, discussed in previous section, the above doubly asymptotic formulae imply interesting scaling relations
with respect to distance and one of the parameters, $\mu$ or $J$, of the correlation function $G(r^{\prime},r^{\prime})$, or of some of its factors. Specifically, in the corresponding doubly asymptotic regions, formulae (\ref{G_diag_asympt_2d+}) and (\ref{diag_asympt 2d+ mu=0}) imply that
\begin{equation}
|G(r^{\prime},r^{\prime})| \approx |\mu| d_{J}(|\mu| r^{\prime}),
\label{scal mu r r}
\end{equation}
from (\ref{G_diag_asympt_2d+}) and (\ref{diag_asympt 2d+ mu=2})  we obtain
\begin{equation}
|G(r^{\prime},r^{\prime})| \approx |J| d_{|\mu|=2}(\sqrt{|J|} r^{\prime}).
\label{scal J r r mu=2}
\end{equation}
Finally, from (\ref{G_diag_asympt_2d+}) and (\ref{diag_asympt 2d+ J=0}) we derive
\begin{equation}
{\cal{C}}{\cal{D}} \approx |J|^{3/2} d_{\mu}(|J| r^{\prime}),
\label{scal J r r}
\end{equation}
with ${\cal{C}}$, ${\cal{D}}$ defined as in section \ref{1d}, and where $d_J$, $d_{\mu}$ and $d_{|\mu|=1}$ stand for some functions.
We note that the scaling formulae (\ref{scal mu r r}), (\ref{scal J r r mu=2}) and (\ref{scal J r r}) immediately imply that in the corresponding critical neighborhoods the critical indices $\nu_{\text{diag}}$ assume the values $1$, $1/2$ and $1$, respectively.
Remarkably, similarly to the one-dimensional case numerical calculations  show that the scaling relations (\ref{scal mu r r}), (\ref{scal J r r mu=2}) and (\ref{scal J r r}), derived only for sufficiently large distances, hold in the whole range of distances. Specifically, scaling property
(\ref{scal mu r r}) is demonstrated in Figs. \ref{cf_J01_ry1rx}, \ref{cf_J1_ry1rx}, then scaling property (\ref{scal J r r mu=2}) in Fig.~\ref{cf_mu2_ry1rx}, and finally scaling property (\ref{scal J r r}) in Figs.~\ref{cf_mu01_ry1rx}, \ref{cf_mu1_ry1rx}.

{\bf (ii) offdiagonal directions}: $r_1 {\neq} r_2$ \\

\begin{figure}
\centering
\begin{subfigure}[b]{0.48\textwidth}
\centering
\includegraphics[width=8cm,clip=on]{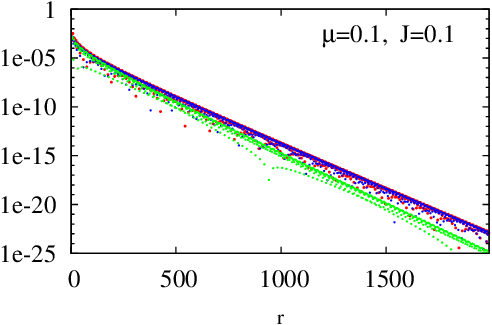}
\end{subfigure}
~
\begin{subfigure}[b]{0.48\textwidth}
\centering
\includegraphics[width=8cm,clip=on]{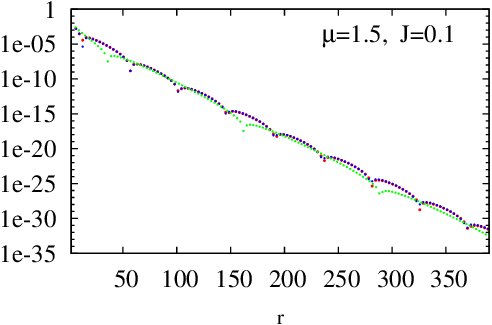}
\end{subfigure}
\caption{\label{cf_mu01_J01_new}\label{cf_mu15_J01_new}(Color online) $d{=}2$, the symmetric model. Plots of $|G(r_1,r_2)|$ versus distance in lattice direction $n{=}3$ -- red points, its asymptotic approximation (formula (\ref{G offdiag_asympt_dist_2d+})) -- blue points, and asymptotic approximation to $|G(r_1,r_2)|$ versus distance in axial direction $r_2{=}0$ -- green  points. All the plots are in logarithmic scale. By (\ref{2offdiag_asympt 2d+ J=0}), in left panel $\xi_2{<}0$ while in right one $\xi_2{>}0$. Consequently the plots of $|G(r_1,r_2)|$ for $n{=}3$ and for $n{=}\infty$ are closer in right panel than in left one. The values of $|G(r_1,r_2)|$ obtained from asymptotic formula (\ref{G offdiag_asympt_dist_2d+}) almost coincide with numerically exact values of $|G(r_1,r_2)|$, already at distances of the order of 100 lattice constants.}
\end{figure}

This case is a bit more involved than the diagonal case; the presentation follows that of previous subsection.
The correlation function $G(r_1,r_2)$, given by (\ref{corr_G 2}), assumes the form
\begin{eqnarray}
G(r_1,r_2)= - \frac{1}{2\pi^2}\int_{0 \leq k_1,k_2 \leq \pi}dk_1 dk_2 \cos(r_1k_1)\cos(r_2k_2)
\frac{\varepsilon_{(k_1,k_2)}}{E_{(k_1,k_2)}},
\label{G offdiag 2d+}
\end{eqnarray}
where $\varepsilon_{(k_1,k_2)}$ and $E_{(k_1,k_2)}$  are defined by (\ref{eps E 2d+}).
In the stripe $|\mu| \leq 2$ of the $(\mu,J)$--plane, but excluding the $\mu{=}0$ and $J{=}0$ lines,
the large-distance $r{=}\sqrt{r_1^2 + r_2^2}$ asymptotic behavior of $G(r_1,r_2)$ reads
\begin{eqnarray}
G(r_1,r_2) \approx
- \frac{{\cal{C}}_{{\bs r}}}{2\pi} \left( \frac{\mu^2 J^2}{1+J^2} \right)^{1/4}
\frac{\exp(-r_1/\xi_1 - r_2^2/(\xi_2 r_1))}{r_1} \cos(\theta_1 r_1 + \theta_2 r_2^2/r_1 + \phi),
\label{G offdiag_asympt 2d+}
\end{eqnarray}
provided the points $(r_1,r_2)$ are located between the ray  $r_1/r_2 \equiv n$, with $n$ being a sufficiently
large rational (in fact it is enough that $n \geq 3$, see Fig.~\ref{cf_mu01_J01_new}), and the $r_2{=}0$-axis. The function of
${\bs r}$, ${\cal{C}}_{{\bs r}}$, is defined as follows
\begin{eqnarray}
{\cal{C}}_{{\bs r}} = \left\{ \begin{array}{ll}
1, & \textrm{if $\mu{>}0$},\\
(-1)^{(r_1+r_2+1)}, & \textrm{if  $\mu{<}0$}.
\end{array} \right.
\label{Cr}
\end{eqnarray}
The numerous constants (independent of coordinates $r_1$ and $r_2$) in (\ref{G offdiag_asympt 2d+}) are given
as follows:
\begin{equation}
\frac{1}{\xi_1}=2\ln\left(a_{+} + \sqrt{a_{+}^2-1}\right), \qquad
\theta_1=2\arccos a_{-}, \qquad
\phi = -\frac{\pi}{4}+\arctan \frac{1}{|J|},
\label{2d+_offdiag_xi1_theta1_phi}
\end{equation}

\begin{eqnarray}
a_{\pm} = \frac{1}{2}\left(\sqrt{(u + 1)^2+v^2} \pm \sqrt{(u-1)^2+v^2}\right),
\label{2d+_offdiag_a}
\end{eqnarray}

\begin{equation}
|u + iv| = \sqrt{\frac{|\mu|}{2\sqrt{1+J^2}}}\,, \qquad \arg(u+iv) = \frac{1}{2}\arctan |J|,
\label{2d+_offdiag_u_v}
\end{equation}

\begin{equation}
\left|\theta_2 - i\frac{1}{\xi_2} \right| =
\sqrt{\frac{|\mu|}{2(1+J^2)}}\left[\left(1-\frac{|\mu|}{2(1+J^2)}\right)^2 +
\left(\frac{\mu J}{2(1+J^2)}\right)^2\right]^{1/4},
\label{2d+_offdiag_theta2_xi2_mod}
\end{equation}

\begin{equation}
\arg \left(\theta_2 - i\frac{1}{\xi_2}\right) =
\frac{1}{2}\arctan |J| - \frac{1}{2}\arctan\frac{|\mu J|}{2(1+J^2)-|\mu|}\,.
\label{2d+_offdiag_theta2_xi2_arg}
\end{equation}

In particular, if the points $(r_1,r_2)$ become remote from the origin along a ray $r_1/r_2{=} n {=} const$, then
in terms of the distance $r$ ($r{=}r_1\sqrt{(1+n^2)/n^2}$),
the asymptotic formula (\ref{G offdiag_asympt 2d+}) can be rewritten as
\begin{equation}
G(r_1,r_2) \approx
-  \frac{{\cal{C}}_{{\bs r}}}{2\pi} \left( \frac{\mu^2 J^2}{1+J^2} \right)^{1/4} \left( \frac{1+n^2}{n^2} \right)^{1/2}
\frac{\exp \left( -r/\xi_{\text{offdiag}}^{(+)} \right)}{r} \cos (r \theta_{\text{offdiag}} + \phi ),
\label{G offdiag_asympt_dist_2d+}
\end{equation}
defining  $\xi_{\text{offdiag}}^{(+)}$ -- the correlation length in an offdiagonal direction specified by $n$,
\begin{equation}
\frac{1}{\xi_{\text{offdiag}}^{(+)}} =
\left( \frac{n^2}{1+n^2} \right)^{1/2} \left( \frac{1}{\xi_1} + \frac{1}{n^2} \frac{1}{\xi_2} \right),
\label{xi_offdiag_2d+}
\end{equation}
and $\theta_{\text{offdiag}}$,
\begin{equation}
\theta_{\text{offdiag}} = \left( \frac{n^2}{1+n^2} \right)^{1/2} \left( \theta_1 + \frac{1}{n^2} \theta_2 \right).
\label{theta_offdiag_2d+}
\end{equation}

\begin{figure}
\begin{center}
\includegraphics[width=15cm,clip=on]{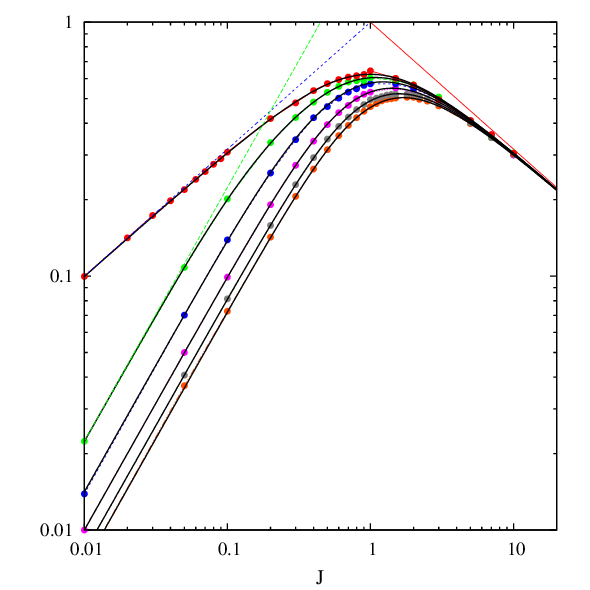}
\caption{\label{offdiag_2d+ xi} (Color online) $d{=}2$, the symmetric model and an axial direction. Plots of $(\sqrt{\mu}\xi_{\text{offdiag}}^{(+)})^{-1}$ versus $J$ in doubly logarithmic scale, in an axial direction ($r_1{=}0$ or $r_2{=}0$), for different values of $\mu$.
Color balls represent the values extracted numerically from the large-distance behavior of $G(r_1,r_2)$;
from top to bottom: $\mu{=}2.0$ -- red, $\mu{=}1.8$ -- green, $\mu{=}1.5$ -- blue, $\mu{=}1.0$ -- magenta, $\mu{=}0.5$ -- gray and $\mu{=}0.1$ -- orange.
Black-continuous lines  are obtained from formula (\ref{xi_offdiag_2d+}).
The dashed lines are the asymptotic lines as $J \to 0$. For $\mu{=}2$, it is the blue-dashed line, which is the plot of $\sqrt{J}$. For $0{<}\mu{<}2$, the asymptotic lines are the plots of $J/\sqrt{2-\mu}$;
green-dashed line is a representative plot for $\mu{=}1.8$.
The red-continuous line, which is the plot of $J^{-1/2}$, is the asymptotic line as $J \to \infty$, for any
$0{<}\mu{\leq} 2$.}
\end{center}
\end{figure}

\begin{figure}
\begin{center}
\includegraphics[width=15cm,clip=on]{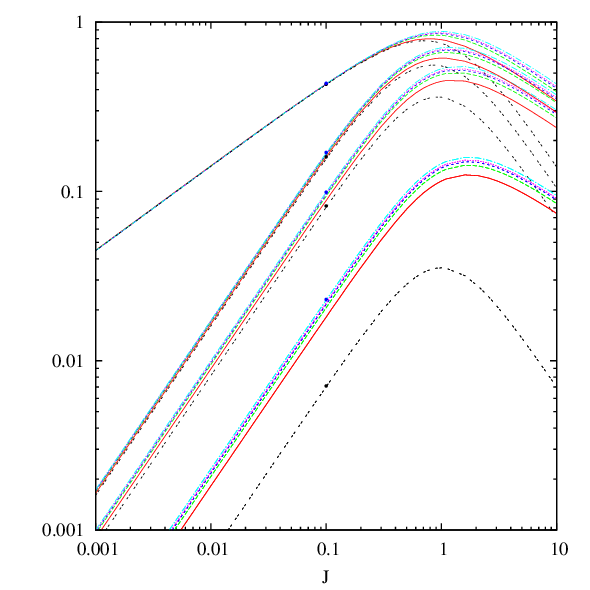}
\caption{\label{1offdiag_2d+_xi_n} (Color online) $d{=}2$, the symmetric model and the diagonal and offdiagonal directions. Plots of $(\xi_{\text{diag}}^{(+)})^{-1}$
and $(\xi_{\text{offdiag}}^{(+)})^{-1}$
versus $J$ in doubly logarithmic scale. There are four groups of lines, each group from bottom to top starts with a black dashed line and ends with a light-blue dashed-dotted line. For each group the value of $\mu$ is: (from bottom to top) $\mu{=}0.1$, $\mu{=}1.0$, $\mu{=}1.5$ and $\mu{=}2.0$. Each group consists of four plots corresponding to different directions $n$ (from bottom to top): $n{=}1$ -- black dashed line, $n{=}2$ -- red continuous line, $n{=}3$ -- green dashed line, $n{=}4$ -- blue dashed line, $n{=}5$ -- magenta dotted line and $n{=}\infty$ -- light-blue dashed-dotted line. The bullets represent $(\xi_{\text{diag}}^{(+)})^{-1}$ -- black ones,
and  $(\xi_{\text{offdiag}}^{(+)})^{-1}$ -- blue ones, extracted numerically from the large-distance behavior of $G(r_1,r_2)$. See also a magnification of the plots around $J{=}0.1$, Fig.~\ref{2offdiag_2d+_xi_n}.}
\end{center}
\end{figure}

\begin{figure}
\begin{center}
\includegraphics[width=15cm,clip=on]{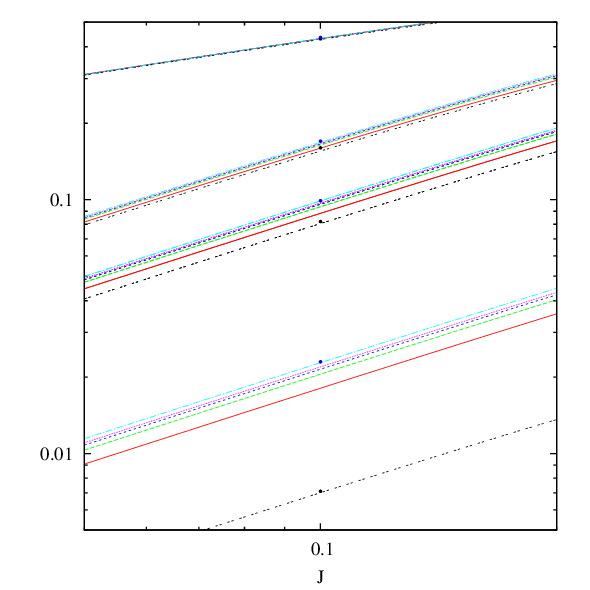}
\caption{\label{2offdiag_2d+_xi_n} (Color online) $d{=}2$, the symmetric model and the diagonal and offdiagonal directions. This is a magnification of Fig.~\ref{1offdiag_2d+_xi_n}, for details see the caption of that figure.}
\end{center}
\end{figure}

The above formula for the correlation length $\xi_{\text{offdiag}}^{(+)}$  is in excellent agreement with the correlation length determined numerically, which is demonstrated in Fig.~\ref{offdiag_2d+ xi}, where plots of $\xi_{\text{offdiag}}^{(+)}$, in an axial direction, against $J$, for a number of $\mu$-values, are displayed.
To reveal the dependence on direction $n$, in Figs.~\ref{1offdiag_2d+_xi_n} and \ref{2offdiag_2d+_xi_n}  we display similar plots, but now, for a fixed $\mu$  we show plots for a number of directions $n$.

Naturally, the large-distance asymptotic formula (\ref{G offdiag_asympt 2d+}) can be simplified in doubly asymptotic regions; we provide only simplified expressions for $\xi_1$, $\xi_2$, $\theta_1$ and $\theta_2$.
Specifically, in doubly asymptotic regions, where $(\mu,J)$-points approach along a $\mu$-path a point belonging to any one of the two half lines of quantum critical points, $\mu{=}0$ and $J{\neq} 0$, formulae
(\ref{2d+_offdiag_xi1_theta1_phi})-(\ref{2d+_offdiag_theta2_xi2_arg}) imply that
\begin{equation}
\frac{1}{\xi_1} \approx \sqrt{\frac{2|\mu|}{\sqrt{1+J^2}}}\sin\left(\frac{1}{2}\arctan |J| \right) = - \frac{2}{\xi_2}, \qquad
\theta_1 \approx \pi- 2 \theta_2,
\label{1offdiag_asympt 2d+ mu=0}
\end{equation}
with
\begin{equation}
\theta_2 \approx \sqrt{\frac{|\mu|}{2\sqrt{1+J^2}}}\cos\left(\frac{1}{2}\arctan |J| \right),
\label{2offdiag_asympt 2d+ mu=0}
\end{equation}
then if $(\mu,J)$-points approach along the $J$-path one of the two end critical points, $|\mu|{=}2$ and $J{=}0$,
\begin{equation}
\frac{1}{\xi_1} \approx \sqrt{2|J|} = \frac{2}{\xi_2}, \qquad
\theta_1 \approx \sqrt{2|J|} = 2\theta_2,
\label{offdiag_asympt 2d+ mu=2}
\end{equation}
and finally if $(\mu,J)$-points approach along a $J$-path any point of the two line segments of quantum critical points, $J{=}0$ and $0< |\mu| <2$,
\begin{eqnarray}
\frac{1}{\xi_1} \approx |J|\sqrt{\frac{|\mu|}{2-|\mu|}}, \qquad
\theta_1 \approx \pi - 2\arcsin\sqrt{\frac{|\mu|}{2}} +\frac{1}{2}\sqrt{\frac{|\mu|}{2-|\mu|}}\frac{3-|\mu|}{2-|\mu|}J^2,
\label{1offdiag_asympt 2d+ J=0}
\end{eqnarray}

\begin{eqnarray}
\frac{1}{\xi_2} \approx \frac{|\mu|-1}{2} \frac{1}{\xi_1}, \qquad
\theta_2 \approx  \frac{1}{2}\sqrt{{|\mu|}{(2-|\mu|)}}.
\label{2offdiag_asympt 2d+ J=0}
\end{eqnarray}

Here, it is worth to add a few comments concerning the large-distance asymptotic behaviour of $G(r_1,r_2)$ in offdiagonal directions.
Clearly, $\lim_{n \to \infty}\xi_{\text{offdiag}}^{(+)} {=} \xi_1$, that is $\xi_1$ is the correlation length in the axial direction $r_2{=}0$, and as such it is necessarily positive. In distinction to  $\xi_1$, the quantity $\xi_2$ cannot be interpreted as a correlation length, since it may even be negative as formulae (\ref{1offdiag_asympt 2d+ mu=0}) and (\ref{2offdiag_asympt 2d+ J=0}) show. The quantity $\xi_2$ is rather a correction to the axial correlation length as direction $n$ deviates from the axial one. Irrespectively of the sign of $\xi_2$, $\xi_{\text{offdiag}}^{(+)}$ is a decreasing function of parameter $n$
(clarify Figs.~\ref{1offdiag_2d+_xi_n} and \ref{2offdiag_2d+_xi_n}),  that is $\xi_{\text{offdiag}}^{(+)}>\xi_1$, for any direction $n$.
 Consequently, $G(r_1,r_2)$ decreases with distance in direction $n$ slower than in the axial one, see Fig.~\ref{cf_mu01_J01_new}.  However, the sign of $\xi_2$ determines how close the plots of $G(r_1,r_2)$  in direction $n$ and in the axial one are. For sufficiently small $J$, (\ref{2offdiag_asympt 2d+ J=0}) implies that $\xi_2{>}0$ ($\xi_2{<}0$) if $|\mu|{>}1$
($|\mu|{<}1$). Therefore, in an offdiagonal direction $n$ the following inequality holds:
$(\xi_{\text{offdiag}}^{(+)}/\xi_1)|_{|\mu|<1}{>}(\xi_{\text{offdiag}}^{(+)}/\xi_1)|_{|\mu|>1}$. In Fig.~\ref{cf_mu01_J01_new} it is well visible that the plots of $G(r_1,r_2)$  for $n{=}3$ and $n{=}\infty$ are more close for $\mu{=}1.5$, than for $\mu=0.1$.
Fig.~\ref{cf_mu01_J01_new} demonstrates also a remarkable agreement between $G(r_1,r_2)$ and its asymptotic approximation (\ref{G offdiag_asympt 2d+}), already at not very large distances as distances of the order of hundreds of lattice constants, and already for such a small $n$ as $n{=}3$.

\begin{figure}
\centering
\begin{subfigure}[b]{0.48\textwidth}
\centering
\includegraphics[width=8cm,clip=on]{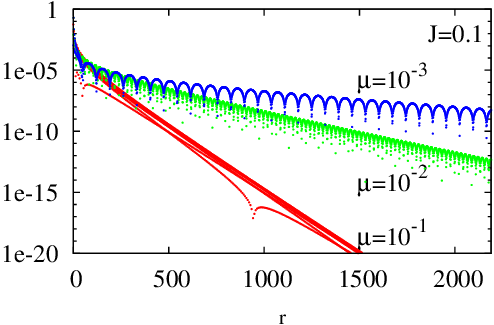}
\end{subfigure}
~
\begin{subfigure}[b]{0.48\textwidth}
\centering
\includegraphics[width=8cm,clip=on]{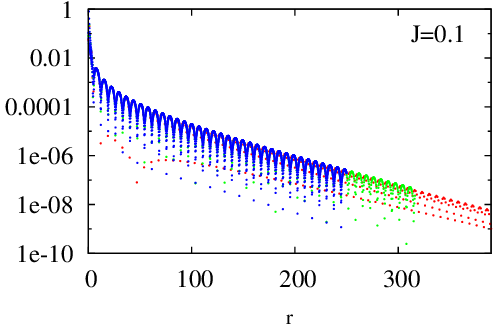}
\end{subfigure}
\caption{\label{cf_J01_ry0}\label{cf_J01_ry0_sc}(Color online) $d{=}2$, the symmetric model and an offdiagonal direction. Left panel, plots of $|G(r,0)|$ in logarithmic scale
for $J{=}0.1$ and three values of $\mu$: $\mu=0.1$ -- red line, $\mu{=}0.01$ -- green line, $\mu{=}0.001$ -- blue line.  Right panel, the plot for $\mu{=}0.1$ is repeated, while the remaining ones are scaled according to formula (\ref{scal mu offdiag}). Clearly, all the plots merge into one plot in the whole range of distances.}
\end{figure}

\begin{figure}
\centering
\begin{subfigure}[b]{0.48\textwidth}
\centering
\includegraphics[width=8cm,clip=on]{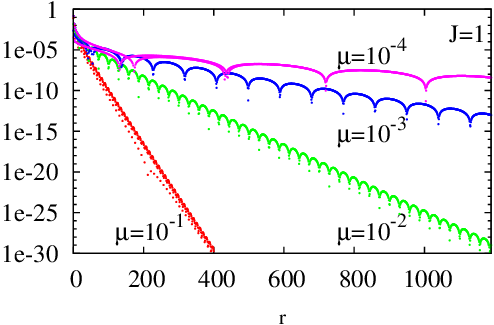}
\end{subfigure}
~
\begin{subfigure}[b]{0.48\textwidth}
\centering
\includegraphics[width=8cm,clip=on]{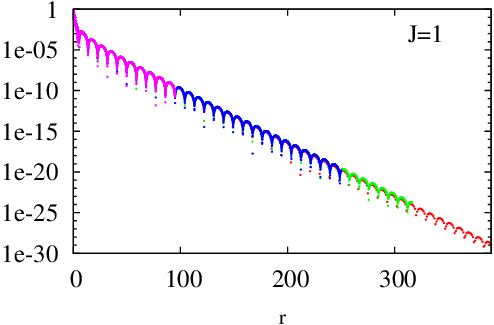}
\end{subfigure}
\caption{\label{cf_J1_ry0}\label{cf_J1_ry0_sc}(Color online) $d{=}2$, the symmetric model and an offdiagonal direction. Left panel, plots of $|G(r,0)|$ in logarithmic scale
for $J{=}1$ and four values of $\mu$: $\mu{=}0.1$ -- red line, $\mu{=}0.01$ -- green line, $\mu{=}0.001$ -- blue line, $\mu{=}0.0001$ -- magenta line. Right panel, the plot for $\mu{=}0.1$ is repeated, while the remaining ones are scaled according to formula (\ref{scal mu offdiag}). Clearly, all the plots merge into one plot in the whole range of distances.}
\end{figure}

\begin{figure}
\centering
\begin{subfigure}[b]{0.48\textwidth}
\centering
\includegraphics[width=8cm,clip=on]{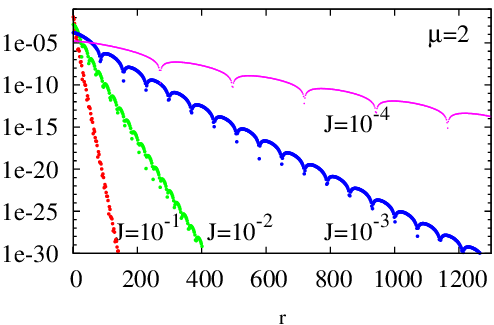}
\end{subfigure}
~
\begin{subfigure}[b]{0.48\textwidth}
\centering
\includegraphics[width=8cm,clip=on]{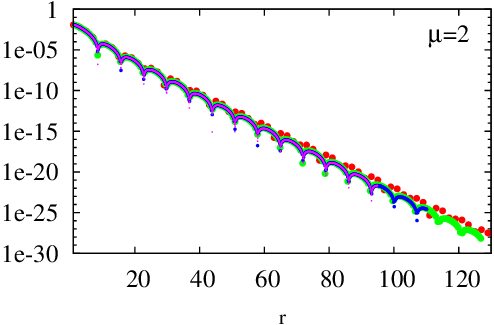}
\end{subfigure}
\caption{\label{cf_mu2_ry0}\label{cf_mu2_ry0_sc}(Color online) $d{=}2$, the symmetric model and an offdiagonal direction. Left panel, plots of $|G(r,0)|$ in logarithmic scale
for $\mu{=}2$ and four values of $J$: $J{=}0.1$ -- red line, $J{=}0.01$ -- green line, $J{=}0.001$ -- blue line, $J{=}0.0001$ -- magenta line. Right panel, the plot for $J{=}0.1$ is repeated, while the remaining ones are scaled according to formula (\ref{scal J_offdiag_mu=2}). Clearly, all the  plots merge into one plot in the whole range of distances. It is well seen that scaling properties hold only sufficiently close to the critical point; $J{=}0.1$ is too large and the corresponding plot does not coincide with the other ones.}
\end{figure}

\begin{figure}
\centering
\begin{subfigure}[b]{0.48\textwidth}
\centering
\includegraphics[width=8cm,clip=on]{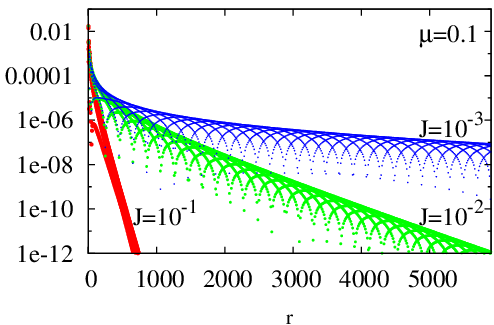}
\end{subfigure}
~
\begin{subfigure}[b]{0.48\textwidth}
\centering
\includegraphics[width=8cm,clip=on]{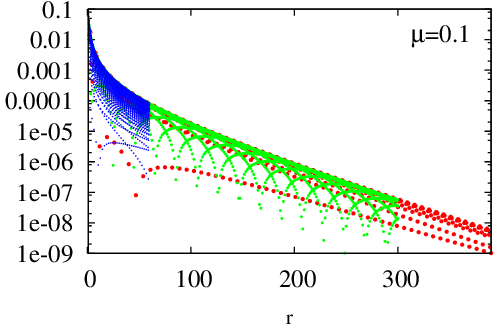}
\end{subfigure}
\caption{\label{cf_mu01_ry0}\label{cf_mu01_ry0_sc}(Color online) $d{=}2$, the symmetric model and an offdiagonal direction. Left panel, plots of $|G(r,0)|$ in logarithmic scale
for $\mu{=}0.1$ and three values of $J$: $J{=}0.1$ -- red line, $J{=}0.01$ -- green line, $J{=}0.001$ -- blue line. Right panel, the plot for $J{=}0.1$ is repeated, while the remaining ones are scaled according to formula (\ref{scal J_offdiag}). Clearly, only the envelops of the plots of $|G(r,0)|$, describing decay of correlations with distance, merge into one plot in the whole range of distances.}
\end{figure}

\begin{figure}
\centering
\begin{subfigure}[b]{0.48\textwidth}
\centering
\includegraphics[width=8cm,clip=on]{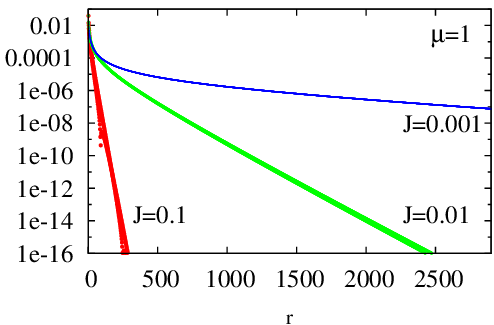}
\end{subfigure}
~
\begin{subfigure}[b]{0.48\textwidth}
\centering
\includegraphics[width=8cm,clip=on]{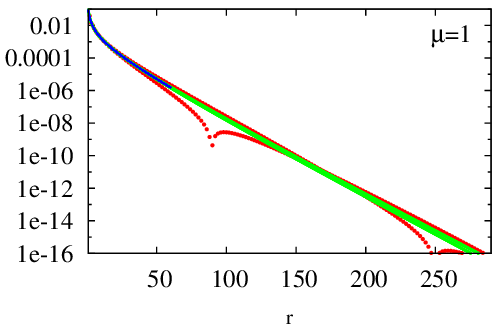}
\end{subfigure}
\caption{\label{cf_mu1_ry0}\label{cf_mu1_ry0_sc}(Color online) $d{=}2$, the symmetric model and an offdiagonal direction. Left panel, plots of $|G(r,0)|$ in logarithmic scale
for $\mu{=}1$ and three values of $J$: $J{=}0.1$ -- red line, $J{=}0.01$ -- green line, $J{=}0.001$ -- blue line. Right panel, the plot for $J{=}0.1$ is repeated, while the remaining ones are scaled according to formula (\ref{scal J_offdiag}). Clearly, only the envelops of the plots of $|G(r,0)|$, describing decay of correlations with distance, merge into one plot in the whole range of distances.}
\end{figure}

As in the previously considered cases of asymptotic behavior of the two-point correlation  function, in one-dimensional model and in the symmetric two-dimensional model and in the diagonal direction, the above doubly asymptotic formulae imply interesting scaling laws with respect to distance and one of the parameters, $\mu$ or $J$, of the correlation function $G(r_1,r_2)$ in offdiagonal directions, or to some of its factors (see section \ref{1d}). Specifically,  in the corresponding doubly asymptotic regions, formulae
(\ref{G offdiag_asympt 2d+}), (\ref{1offdiag_asympt 2d+ mu=0})  and (\ref{2offdiag_asympt 2d+ mu=0})  imply that
\begin{equation}
|G(r_1,r_2)| \approx |\mu| f_{J}(\sqrt{|\mu|} r),
\label{scal mu offdiag}
\end{equation}
while from (\ref{G offdiag_asympt 2d+}) and (\ref{offdiag_asympt 2d+ mu=2}) we derive
\begin{equation}
|G(r_1,r_2)| \approx |J| f_{|\mu|=2}(\sqrt{|J|} r).
\label{scal J_offdiag_mu=2}
\end{equation}
Finally, from (\ref{G offdiag_asympt 2d+}), (\ref{1offdiag_asympt 2d+ J=0}) and (\ref{2offdiag_asympt 2d+ J=0}) we obtain
\begin{equation}
{\cal{C}}{\cal{D}} \approx |J|^{3/2} f_{\mu}(|J| r),
\label{scal J_offdiag}
\end{equation}
with ${\cal{C}}$, ${\cal{D}}$ defined as in section \ref{1d} and $f_{J}$, $f_{|\mu|=2}$, $f_{\mu}$ being some functions.
Let us note again that the scaling formulae (\ref{scal mu offdiag}), (\ref{scal J_offdiag_mu=2}) and (\ref{scal J_offdiag}) immediately imply that in the corresponding critical neighborhoods the critical indices $\nu_{\text{offdiag}}$ assume the values $1/2$, $1/2$ and $1$, respectively.
It is remarkable that, as numerical calculations show, the scaling relations (\ref{scal mu offdiag}),
(\ref{scal J_offdiag_mu=2}) and (\ref{scal J_offdiag}), derived only for sufficiently large distances, hold in the whole range of distances. Specifically, scaling property (\ref{scal mu offdiag}) is demonstrated in Figs.~\ref{cf_J01_ry0}, \ref{cf_J1_ry0}, then scaling property (\ref{scal J_offdiag_mu=2}) in Fig.~\ref{cf_mu2_ry0},
and finally scaling property (\ref{scal J_offdiag}) in Figs.~\ref{cf_mu01_ry0}, \ref{cf_mu1_ry0}.

\subsection{The antisymmetric model}

It is worth to mention that the Hamiltonian of the antisymmetric model can be obtained as a mean-field approximation to
\begin{equation}
 \sum_{{\bs k},\sigma}\varepsilon_{\bs k} c^{\dagger}_{{\bs k},\sigma} c_{{\bs k},\sigma}
 + J \sum_{{\bs l},i} {\bs S_{{\bs l}} } {\bs S_{{\bs l}+{\bs e}_i} },
\label{spin ham}
\end{equation}
where ${\bs S_{{\bs l}}}$ stands for the spin operator of an electron at site ${\bs l}$ of the underlying lattice (for the notation see section \ref{models}). The parameters $\Delta_{i}$ of the mean-field Hamiltonian are no longer free parameters but are given implicitly by those solutions
of the equations
\begin{equation}
\Delta_{i} = - \langle  a_{{\bs l},\uparrow} a_{{\bs l}+{\bs e}_i,\downarrow} -
a_{{\bs l},\downarrow} a_{{\bs l}+{\bs e}_i,\uparrow} \rangle,
\label{delta}
\end{equation}
where the brackets denote the Gibbs average with the mean-field Hamiltonian, that minimize the ground-state energy -- physical solutions.
When the underlying lattice is a square lattice, it turns out that the physical solutions satisfy the condition $\Delta_1{=}-\Delta_2$, which corresponds to the so called $d_{x^2-y^2}$ pairing in theory of $d$-wave superconductivity.

\begin{figure}
\begin{center}
\includegraphics[width=10cm,clip=on]{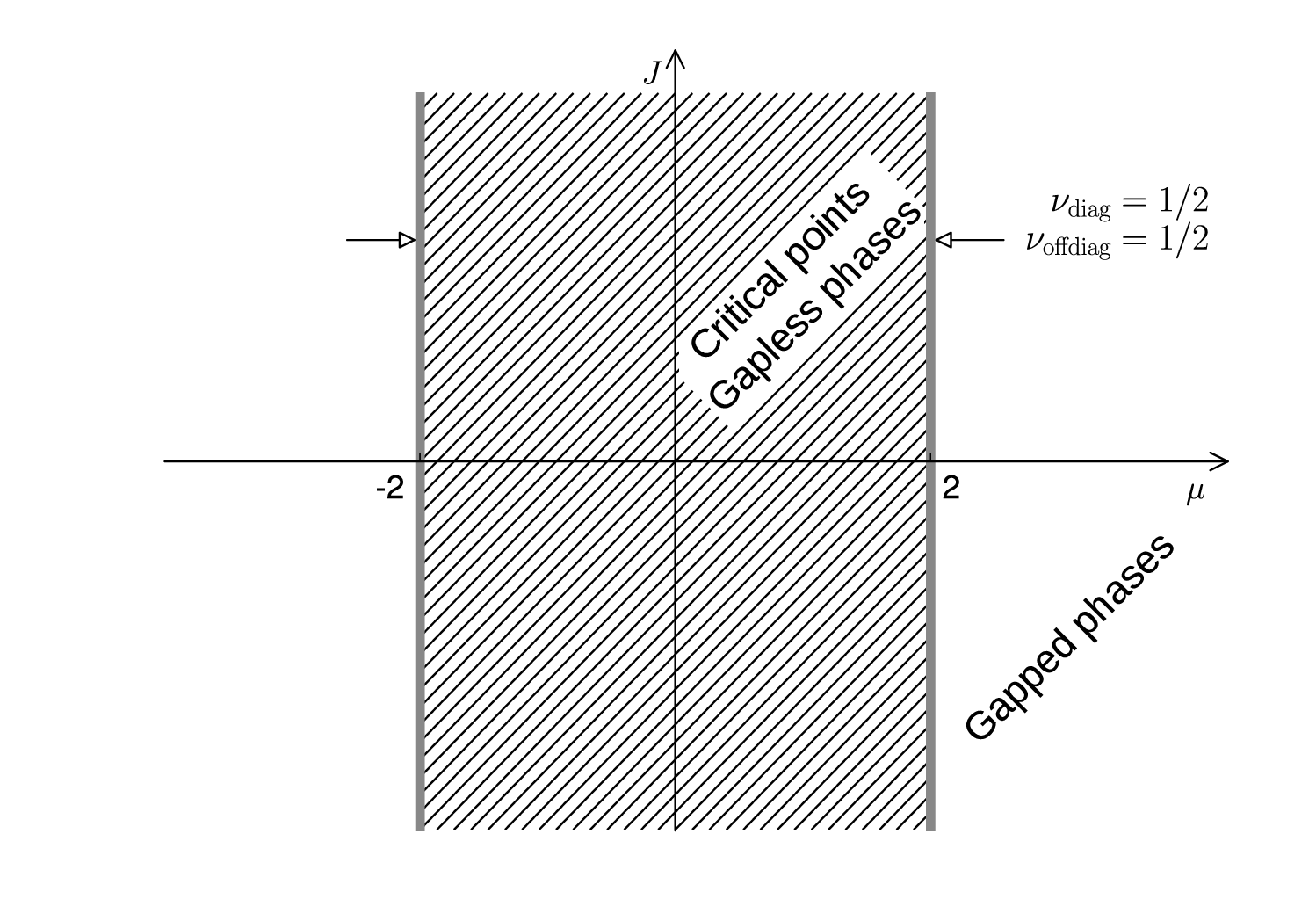}
\caption{\label{diagram2d-} Phase diagram of the antisymmetric two-dimensional system in the $(\mu,J)$-plane. The set of critical points constitutes the stripe  $|\mu|{\leq} 2$.}
\end{center}
\end{figure}

In distinction to the previously considered cases of the one-dimensional model and the symmetric two-dimensional model, where the critical points are located at straight, intersecting  lines, the quantum-critical points of the antisymmetric two-dimensional model fill up the stripe in between the two lines $|\mu|{=}2$, Fig.~\ref{diagram2d-}.

{{\bf (i) the diagonal direction}}: $r_1{=}r_2{=}r^{\prime}$ \\

\begin{figure}
\centering
\begin{subfigure}[b]{0.48\textwidth}
\centering
\includegraphics[width=8cm,clip=on]{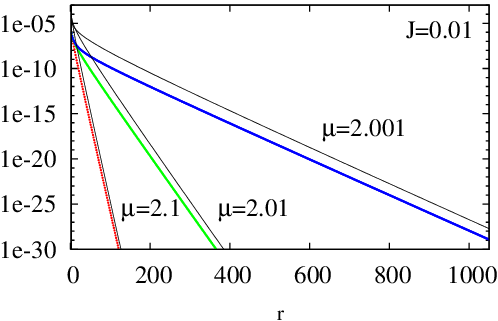}
\end{subfigure}
~
\begin{subfigure}[b]{0.48\textwidth}
\centering
\includegraphics[width=8cm,clip=on]{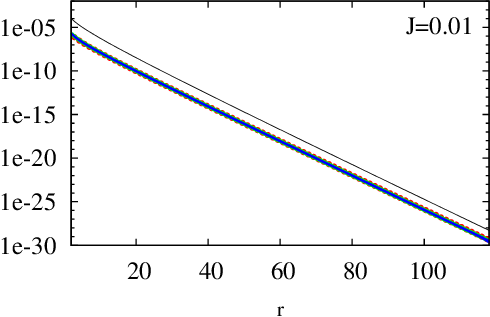}
\end{subfigure}
\caption{\label{cf_J001_ry1rx}\label{cf_J001_ry1rx_sc}(Color online) $d{=}2$, the antisymmetric model and the diagonal direction. Left panel, plots of numerically calculated $|G(r^{\prime},r^{\prime})|$  versus distance $r$, in logarithmic scale, for $J{=}0.01$ and three values of $\mu$: $\mu{=}2.1$ -- red line, $\mu{=}2.01$ -- green line, $\mu{=}2.001$ -- blue line. Plots obtained from doubly asymptotic formulae (\ref{G diag_asympt 2d-}), (\ref{xi diag_2d-}) for the same values of $J$ and $\mu$ -- black lines. Right panel, the plots for $\mu{=}2.1$ are repeated, while the remaining ones are scaled according to formula (\ref{G diag 2d- scal}). All the plots obtained from doubly asymptotic formula
(\ref{G diag_asympt 2d-}) merge into one plot, even for rather small distances, and so do the numerically obtained plots of $|G(r^{\prime},r^{\prime})|$. For $J{=}0.01$ (and smaller values of $J$, not shown) the asymptotic formula (\ref{G diag_asympt 2d-}) does not fit well $|G(r^{\prime},r^{\prime})|$.}
\end{figure}

\begin{figure}
\centering
\begin{subfigure}[b]{0.48\textwidth}
\centering
\includegraphics[width=8cm,clip=on]{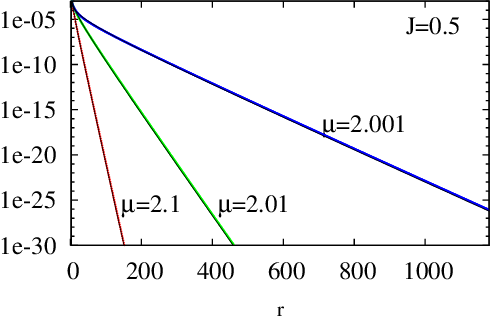}
\end{subfigure}
~
\begin{subfigure}[b]{0.48\textwidth}
\centering
\includegraphics[width=8cm,clip=on]{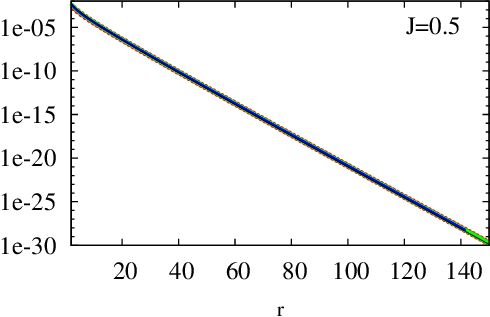}
\end{subfigure}
\caption{\label{cf_J05_ry1rx}\label{cf_J05_ry1rx_sc}(Color online) $d{=}2$, the antisymmetric model and the diagonal direction. Left panel, plots of numerically calculated $|G(r^{\prime},r^{\prime})|$  versus distance $r$, in logarithmic scale,
for $J{=}0.5$ and three values of $\mu$: $\mu{=}2.1$ -- red line, $\mu{=}2.01$ -- green line, $\mu{=}2.001$ -- blue line. Plots obtained from doubly asymptotic formulae (\ref{G diag_asympt 2d-}), (\ref{xi diag_2d-}) for the same values of $J$ and $\mu$ -- black lines. Right panel, the plots for $\mu{=}2.1$ are repeated, while the remaining ones are scaled according to formula (\ref{G diag 2d- scal}). All the plots obtained from doubly asymptotic formula (\ref{G diag_asympt 2d-}) merge into one plot, even for rather small distances, and so do the numerically obtained plots of $|G(r^{\prime},r^{\prime})|$. In distinction to the case $J{=}0.01$, for $J{=}0.5$ (and larger values, not shown) the asymptotic formula
(\ref{G diag_asympt 2d-}) does fit well $|G(r^{\prime},r^{\prime})|$. }
\end{figure}

The correlation function $G(r_1,r_2)$, given by (\ref{corr_G 2}), assumes the form

\begin{eqnarray}
{G}(r^{\prime},r^{\prime})= - \frac{1}{2\pi^2}\int_{0 \leq k_1,k_2 \leq \pi}dk_1 dk_2
\cos(r^{\prime}k_1)\cos(r^{\prime}k_2)
\frac{\varepsilon_{(k_1,k_2)}}{E^{(-)}_{(k_1,k_2)}},
\label{G_diag_2d-}
\end{eqnarray}
with $\varepsilon_{(k_1,k_2)}$ defined by (\ref{eps E 2d+}) and $E^{(-)}_{(k_1,k_2)}$ defined as
\begin{equation}
E^{(-)}_{(k_1,k_2)} = \sqrt{(\cos k_1 + \cos k_2 -\mu)^2 + J^2(\cos k_1 - \cos k_2)^2}.
\label{E 2d-}
\end{equation}
In a doubly asymptotic region, where $(\mu,J)$-points, with $|\mu|{>}2$ and $|J|$ not too close to zero, approach along a $\mu$-path a point  belonging to one of the lines $|\mu|{=}2$, ${G}(r^{\prime},r^{\prime})$ behaves as
\begin{eqnarray}
{G}(r^{\prime},r^{\prime}) \approx - \textrm{sgn} (\mu) \frac{J}{4\pi \xi^{(-)}} \frac{\exp(-r^{\prime}/\xi^{(-)})}{{r^{\prime}}},
\label{G diag_asympt 2d-}
\end{eqnarray}
where
\begin{equation}
\frac{1}{\xi^{(-)}} \approx 2\sqrt{\frac{|\mu| - 2}{1+J^2}},
\label{xi diag_2d-}
\end{equation}
that is the correlation length in the considered doubly asymptotic region is $\xi^{(-)}_{\text{diag}} {=} \sqrt{2}\xi^{(-)}$, and the corresponding $\nu {=} 1/2$. Formula (\ref{G diag_asympt 2d-}) implies the following scaling relation
\begin{equation}
|{G}(r^{\prime},r^{\prime})| \approx ( |\mu| - 2)  h_{\text{diag}}(\sqrt{|\mu| - 2}\,r),
\label{G diag 2d- scal}
\end{equation}
for some function $h_{\text{diag}}$. Interestingly, this scaling relation is valid well beyond the large-distance region as is demonstrated in
Figs.~\ref{cf_J001_ry1rx} and \ref{cf_J05_ry1rx}, provided that $J$ is not too small.

{\bf (ii) offdiagonal directions}

The correlation function $G(r_1,r_2)$, given by (\ref{corr_G 2}), assumes the form
\begin{eqnarray}
G(r_1,r_2) = - \frac{1}{2\pi^2}\int_{0 \leq k_1,k_2 \leq \pi}dk_1 dk_2 \cos(r_1k_1)\cos(r_2k_2)
\frac{\varepsilon_{(k_1,k_2)}}{E^{(-)}_{(k_1,k_2)}},
\label{G offdiag 2d-}
\end{eqnarray}
with $\varepsilon_{(k_1,k_2)}$ defined by (\ref{eps E 2d+}) and $E^{(-)}_{(k_1,k_2)}$  by (\ref{E 2d-}).

In a doubly asymptotic region, where $(\mu,J)$-points, with $|\mu|{>}2$ and $|J|$ not too close to zero, approach along a $\mu$-path a point  belonging to one of the lines $|\mu|{=}2$, the above formula simplifies to

\begin{eqnarray}
G(r_1,r_2) \approx
-  \frac{{\cal{C}}_{{\bs r}}}{2\pi} \left( \frac{(|\mu|-2)|J|}{1+J^2} \right)^{1/4}
\frac{\exp(-r_1/\xi_1^{(-)} - r_2^2/(\xi_2^{(-)} r_1))}{r_1} \cos(\theta_1^{(-)} r_1 +
\theta_2^{(-)} r_2^2/r_1 + \phi^{(-)}),
\label{G offdiag 2d- asympt}
\end{eqnarray}
provided the points $(r_1,r_2)$ are located between a ray  $r_1/r_2 {\equiv} n$, with $n$ being a sufficiently large rational (it is enough that $n {\geq} 3$), and the $r_2{=}0$-axis, and where

\begin{equation}
\frac{1}{\xi_1^{(-)}} \approx \sqrt{\frac{\sqrt{1+J^2}+1}{1+J^2}}\sqrt{|\mu| - 2}, \qquad
\theta_1^{(-)} \approx \sqrt{\frac{\sqrt{1+J^2}-1}{1+J^2}}\sqrt{|\mu| - 2}, \qquad
\phi^{(-)} = \frac{\pi}{4},
\label{xi1 theta1 phi 2d-}
\end{equation}

\begin{equation}
\frac{1}{\xi_2^{(-)}} \approx -\left( \frac{1}{2} - \frac{1}{\sqrt{1+J^2}} \right) \frac{1}{\xi_1^{(-)}}
\label{xi2 2d-}
\end{equation}

\begin{equation}
\theta_2^{(-)} \approx -\left( \frac{1}{2} + \frac{1}{\sqrt{1+J^2}} \right) \theta_1^{(-)}
\label{theta2 2d-}
\end{equation}

\begin{figure}
\begin{center}
\includegraphics[width=15cm,clip=on]{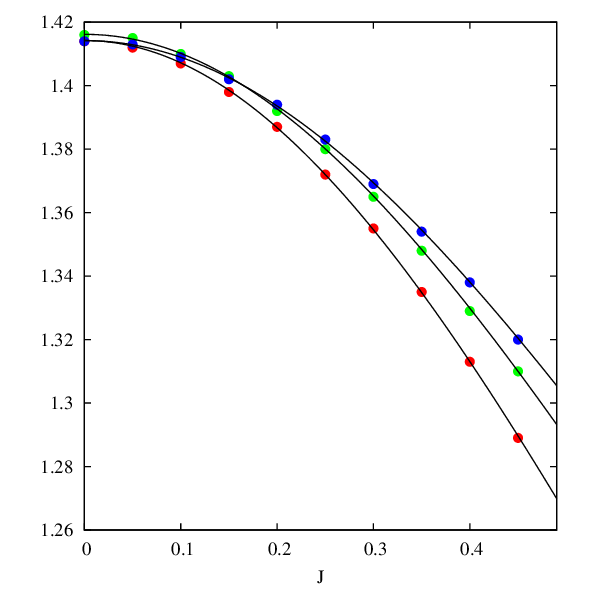}
\caption{\label{xi_AS_J_rynrx_I} (Color online) $d{=}2$, the antisymmetric model and the diagonal and offdiagonal directions. Plots of $(\sqrt{|\mu|-2}\xi_{\text{offdiag}}^{(-)})^{-1}$ and $(\sqrt{|\mu|-2}\xi_{\text{diag}}^{(-)})^{-1}$ versus $J$.
Color balls represent the values extracted numerically from the large-distance behavior of $G(r_1,r_2)$;
from bottom to top: $n{=}1$ -- red, $n{=}3$ -- green, $n{=}\infty$ -- blue.
Black-continuous lines  are obtained from our asymptotic analytic formulae.}
\end{center}
\end{figure}

\begin{figure}
\begin{center}
\includegraphics[width=15cm,clip=on]{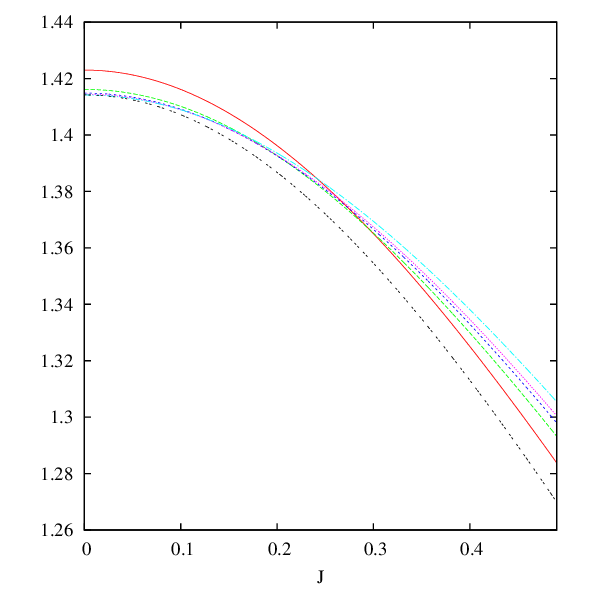}
\caption{\label{xi_AS_J_rynrx_II} (Color online) $d{=}2$, the antisymmetric model and the diagonal and offdiagonal directions. 
Plots of $(\sqrt{|\mu|-2}\xi_{\text{offdiag}}^{(-)})^{-1}$ and $(\sqrt{|\mu|-2}\xi_{\text{diag}}^{(-)})^{-1}$  versus $J$, obtained from our asymptotic analytic formulae, for various directions $n$: (from bottom to top): $n{=}1$ -- black-dashed line, $n{=}2$ -- red-continuous line, $n{=}3$ -- green-dashed line, $n{=}4$ -- blue-dashed line, $n{=}5$ -- magenta-dotted line and $n{=}\infty$ -- light-blue dashed-dotted line. }
\end{center}
\end{figure}

Note that the general forms of the offdiagonal large-distance asymptotic behavior in the symmetric case
(\ref{G offdiag_asympt 2d+}) and in the antisymmetric case (\ref{G offdiag 2d- asympt}) are the same.
Therefore, along a ray $r_1/r_2 {\equiv} n$ with $n$ not too close to $1$, as in the symmetric case we can
rewrite formula (\ref{G offdiag 2d- asympt}) in the form (\ref{G offdiag_asympt_dist_2d+}), simultaneously defining the correlation length in an offdiagonal direction specified by $n$, $\xi_{\text{offdiag}}^{(-)}$ given in terms of $\xi_1^{(-)}$ and $\xi_2^{(-)}$, and $\theta^{(-)}$ given in terms of $\theta_1^{(-)}$ and $\theta_2^{(-)}$. The correlation length $\xi_{\text{offdiag}}^{(-)}$ is given by formula  (\ref{xi_offdiag_2d+}), with $\xi_1$, $\xi_2$ replaced by $\xi_1^{(-)}$, $\xi_2^{(-)}$, respectively, while $\theta^{(-)}$, in an analogous way, is given by (\ref{theta_offdiag_2d+}).
The quality of our approximate formulae for correlation lengths in the diagonal direction (\ref{xi diag_2d-}), and in offdiagonal directions
(\ref{xi_offdiag_2d+}), (\ref{xi1 theta1 phi 2d-}), (\ref{xi2 2d-}) is demonstrated in Fig.~\ref{xi_AS_J_rynrx_I}. 
This figure and Fig.~\ref{xi_AS_J_rynrx_II} reveal also the monotonicity properties with respect to parameter $J$ and direction $n$.

\begin{figure}
\centering
\begin{subfigure}[b]{0.48\textwidth}
\centering
\includegraphics[width=8cm,clip=on]{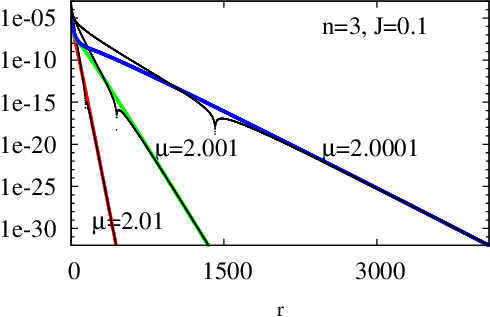}
\end{subfigure}
~
\begin{subfigure}[b]{0.48\textwidth}
\centering
\includegraphics[width=8cm,clip=on]{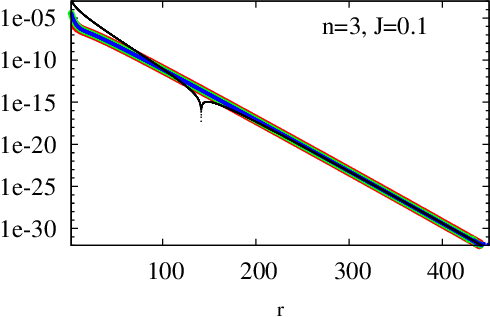}
\end{subfigure}
\caption{\label{cf_J01_ry3rx}\label{cf_J01_ry3rx_sc} (Color online) $d{=}2$, the antisymmetric model and the offdiagonal direction $n{=}3$.
Left panel, plots of numerically calculated $|G(r_1,r_2)|$  versus distance $r$, in logarithmic scale,
for $J{=}0.1$ and three values of $\mu$: $\mu{=}2.01$ -- red line, $\mu{=}2.001$ -- green line, $\mu{=}2.0001$ -- blue line. Plots obtained from doubly asymptotic formulae (\ref{G offdiag 2d- asympt})--(\ref{theta2 2d-}) for the same values of $J$ and $\mu$ -- black lines. Right panel, the plots for $\mu{=}2.01$ are repeated, while the remaining ones are scaled according to formula (\ref{G offdiag 2d- scal}). All the plots obtained from doubly asymptotic formulae (\ref{G offdiag 2d- asympt})--(\ref{theta2 2d-}) merge into one plot, even for rather small distances, and so do the numerically obtained plots of $|G(r_1,r_2)|$. For $J{=}0.01$ (and for smaller values of $J$, not shown) the asymptotic formulae
(\ref{G offdiag 2d- asympt})--(\ref{theta2 2d-}) do not fit well $|G(r_1,r_2)|$. }
\end{figure}

\begin{figure}
\centering
\begin{subfigure}[b]{0.48\textwidth}
\centering
\includegraphics[width=8cm,clip=on]{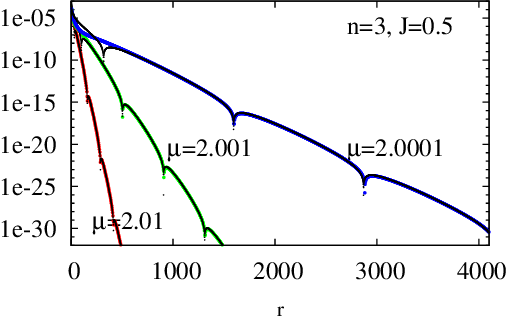}
\end{subfigure}
~
\begin{subfigure}[b]{0.48\textwidth}
\centering
\includegraphics[width=8cm,clip=on]{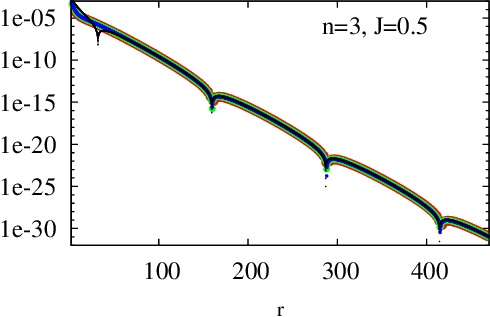}
\end{subfigure}
\caption{\label{cf_J05_ry3rx}\label{cf_J05_ry3rx_sc} (Color online) $d{=}2$, the antisymmetric model and the offdiagonal direction $n{=}3$.
Left panel, plots of numerically calculated $|G(r_1,r_2)|$  versus distance $r$, in logarithmic scale,
for $J{=}0.5$ and three values of $\mu$: $\mu{=}2.01$ -- red line, $\mu{=}2.001$ -- green line, $\mu{=}2.0001$ -- blue line. Plots obtained from doubly asymptotic formulae (\ref{G offdiag 2d- asympt})--(\ref{theta2 2d-}) for the same values of $J$ and $\mu$ -- black lines. Right panel, the plots for $\mu{=}2.01$ are repeated, while the remaining ones are scaled according to formula (\ref{G offdiag 2d- scal}). All the plots obtained from doubly
asymptotic formulae (\ref{G offdiag 2d- asympt})--(\ref{theta2 2d-}) merge into one plot, even for rather small distances, and so do the numerically
obtained plots of $|G(r_1,r_2)|$. In distinction to the case of $J{=}0.1$ and smaller values of $J$, for $J{=}0.5$ (and larger values, not shown) the asymptotic formulae (\ref{G offdiag 2d- asympt})--(\ref{theta2 2d-}) do fit well $|G(r_1,r_2)|$. }
\end{figure}

From formulae (\ref{G offdiag 2d- asympt})--(\ref{theta2 2d-}) one can infer the following scaling relation for $G(r_1,r_2)$:
\begin{equation}
|G(r_1,r_2)| \approx (|\mu| -2) h_{\text{offdiag}}(\sqrt{|\mu| - 2}\,r),
\label{G offdiag 2d- scal}
\end{equation}
for some function $h_{\text{offdiag}}$. This relation, as numerics shows (see Figs.~\ref{cf_J01_ry3rx} and \ref{cf_J05_ry3rx}), holds well beyond the region of large distances.

\section{\label{scaling} Resume of scaling laws and monotonicity properties of correlation lengths}

In previous sections and subsections, we described by means of  analytic formulae and numerical calculations the large-distance asymptotic behavior of the two-point correlation function $G({\bs r})$, for the one-dimensional model and two two-dimensional models, where various lattice directions  and various $\mu$- and $J$--parameter regimes were taken into account. In particular, analytic formulae were obtained and numerical calculation carried out for the associated correlation length. From these results we inferred numerous scaling laws, applying to the correlation length and to $G({\bs r})$, in various asymptotic regimes, as neighborhoods of all types  of critical points  and the asymptotic regime of $|J| \to \infty$ for some $\mu$ separated from zero. The main purpose of this section is to summarize our results concerning scaling laws.

We start with the correlation length.
Analytic formulae for correlation lengths and their asymptotic behaviors in vicinities of all types of critical points exhibited by the studied models are given in:
(\ref{xi theta phi}), (\ref{a}), (\ref{asympt mu=0}), (\ref{asympt mu=1}) and (\ref{asympt J=0}) -- for the one-dimensional system,
in (\ref{2d+_diag_xi theta phi}), (\ref{2d+_diag_a}), (\ref{diag_asympt 2d+ mu=0}), (\ref{diag_asympt 2d+ mu=2}) and (\ref{diag_asympt 2d+ J=0}) -- for the symmetric two-dimensional system in the diagonal direction,
in (\ref{xi_offdiag_2d+}), (\ref{1offdiag_asympt 2d+ mu=0}), (\ref{offdiag_asympt 2d+ mu=2}), (\ref{1offdiag_asympt 2d+ J=0}), and (\ref{2offdiag_asympt 2d+ J=0}) -- for the symmetric two-dimensional system in offdiagonal directions,
in (\ref{xi diag_2d-}) -- for the antisymmetric system in the diagonal direction, and in  (\ref{xi1 theta1 phi 2d-}) and (\ref{xi2 2d-}) -- for the antisymmetric system in offdiagonal directions.
Results of numerical calculations and the plots made using the mentioned formulae, of the correlation length versus $J$ for a number of $\mu$-values are displayed in Fig.~\ref{1d xi} -- for the one-dimensional system, in Fig.~\ref{diag_2d+ xi} -- for the symmetric two-dimensional system in the diagonal direction, and in Fig.~\ref{offdiag_2d+ xi} -- for the symmetric two-dimensional system in offdiagonal directions.

Consider first the asymptotic regime of sufficiently large $|J|$ with fixed $\mu$ separated from zero.
Let $\nu^{(\mu)}$ be the critical exponent for critical points located at $J$-axis (approached along a $\mu$-path).
In the mentioned asymptotic regime the correlation length satisfies the following scaling laws:
\begin{equation}
\left( \delta_l^{\nu_l^{(\mu)}} \xi_l \right)^{-1} = C_l/ |J|^{\alpha_l},\,\,\, \textrm{for sufficiently large $|J|$.}
\label{xi large J}
\end{equation}
The above formula refers to 5 cases labeled by the index $l$: the one-dimensional model ($d{=}1$), the symmetric model (S) and the diagonal spatial direction ($n{=}1$) or an offdiagonal one ($n{\neq}1$), finally the antisymmetric model (AS) and the diagonal or an offdiagonal spatial direction.
The parameter $\delta$ measures the distance to a critical point along a $\mu$-path and depends only on the type of model considered:
\begin{eqnarray}
\delta = \left\{ \begin{array}{ll}
|\mu|, & \textrm{for $d{=}1$,}  \\
|\mu|, & \textrm{for S,}  \\
|\mu|-2, & \textrm{for AS.} \\
\end{array} \right.
\label{delta}
\end{eqnarray}
From the phase diagrams, Fig.~\ref{diagram1d}, Fig.~\ref{diagram2d+} and Fig.~\ref{diagram2d-}, one can read off the values of the critical index $\nu^{(\mu)}$ in each case:
\begin{eqnarray}
\nu^{(\mu)} = \left\{ \begin{array}{lll}
1, & \textrm{for $d{=}1$,}  &  \\
1, & \textrm{for  S,}   & n{=}1, \\
1/2, & \textrm{for S,}  & n {\neq} 1,\\
1/2, & \textrm{for AS,} &  \textrm{any $n$.}\\
\end{array} \right.
\label{numu}
\end{eqnarray}
Then, the exponent $\alpha$ and the coefficient $C$ are given as follows:
\begin{eqnarray}
\alpha = \left\{ \begin{array}{lll}
1, & \textrm{for $d{=}1$,}     &   \\
1, & \textrm{for S or AS,}   & n {=} 1, \\
1/2, & \textrm{for S or AS,} & n {\neq} 1.\\
\end{array} \right.
\label{alpha}
\end{eqnarray}

\begin{eqnarray}
C = \left\{ \begin{array}{lll}
1, & \textrm{for $d{=}1$,} &   \\
\sqrt{\frac{n^2}{1+n^2}}\left( 1-\frac{1}{2n^2} \right), & \textrm{for S or AS,}  & n {\neq} 1, \\
2^{-1/2}, & \textrm{for S,} &  n {=} 1,\\
2^{1/2}, & \textrm{for AS,} &  n {=} 1.\\
\end{array} \right.
\label{C}
\end{eqnarray}
The scaling law (\ref{xi large J}) is well illustrated in Figs.~\ref{1d xi}, \ref{diag_2d+ xi} and \ref{offdiag_2d+ xi}.

Now, consider neighborhoods of critical points located at $\mu$-axis, that is the regime of sufficiently small $|J|$ and fixed $\mu$ separated from zero.
For $d{=}1$-model and the symmetric model, let $\nu^{(J)}$ be the critical exponent for critical points located at $\mu$-axis (approached along a $J$-path). In the specified asymptotic regime the scaling law for the correlation length reads:
\begin{equation}
\left( |\mu|^{\nu_l^{(\mu)}} \xi_l \right)^{-1} = D_l |J|^{\nu_l^{(J)}},\,\,\, \textrm{for sufficiently small $|J|$.}
\label{xi small J}
\end{equation}
Here the index $l$ labels 6 cases specified as follows: $d{=}1$-model and the type of critical point ($|\mu|{\neq} 1$ or $|\mu|{=}1$), the symmetric model and the type of critical point ($|\mu|{\neq} 2$ or $|\mu|{=}2$) and the type of lattice direction.
From the phase diagrams, Fig.~\ref{diagram1d} and Fig.~\ref{diagram2d+}, one can read off the values of the critical exponent $\nu^{(J)}$:
\begin{eqnarray}
\nu^{(J)} = \left\{ \begin{array}{llll}
1, & \textrm{for $d{=}1$,} & |\mu|\neq 1, &  \\
1/2, & \textrm{for  $d{=}1$,} & |\mu|= 1, &  \\
1, & \textrm{for S,} & |\mu|\neq 2, &  \textrm{any $n$,}\\
1/2, & \textrm{for S,} & |\mu| = 2, &  \textrm{any $n$.}\\
\end{array} \right.
\label{nuJ}
\end{eqnarray}
The coefficient $D$ is given as follows
\begin{eqnarray}
D = \left\{ \begin{array}{llll}
(1 - \mu^2)^{-1/2}, & \textrm{for $d{=}1$,} & |\mu| {\neq} 1, &  \\
1, & \textrm{for  $d{=}1$,} & |\mu|{=}1, &  \\
\sqrt{\frac{n^2}{1+n^2}}\left(1 + \frac{|\mu|-1}{2n^2} \right)(2-|\mu|)^{-1/2}, & \textrm{for S,} & |\mu|{\neq} 2, &
n{\neq}1,\\
\sqrt{\frac{n^2}{1+n^2}}\left( 1+\frac{1}{2n^2} \right), & \textrm{for S,} & |\mu|{=}2, &  n{\neq}1, \\
\sqrt{2}, & \textrm{for S,} & |\mu|{=}2, &  n{=}1,\\
\sqrt{\frac{2}{4-\mu^2}}, & \textrm{for S,}  & |\mu|{\neq}2, &  n{=}1.\\
\end{array} \right.
\label{D}
\end{eqnarray}

Besides the above formulated scaling laws, we add here a few general observations concerning monotonicity properties of the correlation length.
In the case of the one-dimensional model and for any lattice direction of the two-dimensional symmetric model,
the plots of the correlation length versus $J$ for a number of $\mu$-values, Figs.~\ref{1d xi}, \ref{diag_2d+ xi}, \ref{offdiag_2d+ xi}, and \ref{1offdiag_2d+_xi_n} show that for $J{>}0$ and $0{<}\mu{\leq}2$ the inverse correlation length as a function of $J$ is
nonmonotonic, it has a maximum for $J{\approx} 1$.
However, as Figs.~\ref{1d xi}, \ref{diag_2d+ xi}, \ref{offdiag_2d+ xi}, \ref{1offdiag_2d+_xi_n}, and \ref{2offdiag_2d+_xi_n} reveal, the inverse correlation length is monotonic with respect to $\mu$ and $n$. Namely, for fixed $J{>}0$ and $n$, $\xi^{-1}$ (one-dimensional model), $\xi_{\text{diag}}^{-1}$, and $\xi_{\text{offdiag}}^{-1}$ in a  direction $n$, are strictly increasing functions of $\mu$. Then, for fixed $J{>}0$ and fixed $\mu$, $0 {<} \mu {\leq} 2$, $\xi_{\text{offdiag}}^{-1}$ is an increasing function of $n$. We note also that, for $J{<}1$ the dependence on $n$ becomes weaker as $\mu$ increases; if $\mu{=}2$  (the end critical point), $\xi_{\text{offdiag}}^{-1}$ is practically constant in $n$. Such an approximate independence of $n$ holds also for fixed $J$ and sufficiently small $\mu$, as readily follows from
(\ref{xi_offdiag_2d+}) and (\ref{1offdiag_asympt 2d+ mu=0}).

On the other hand, Fig.~\ref{xi_AS_J_rynrx_II} reveals that in the case of the two-dimensional antisymmetric model, in distinction to the previously discussed models, the inverse correlation length is monotonic in $J$ -- a strictly decreasing function of parameter $J$. As a function of spatial direction $n$, the inverse correlation length is strictly increasing for sufficiently large $J$ (roughly for $J{>}1/4$). For $J{<}1/4$, the dependence on $n$ is not monotonic; Fig.~\ref{xi_AS_J_rynrx_II} suggests that the inverse correlation length for $n{=}2$ is an upper bound for
inverse correlation lengths in all the other directions. Whatever $J$ is, the inverse correlation length in the diagonal direction is a lower bound for inverse correlation lengths in all the other directions.

Finally, we turn to scaling laws for the two-point correlation function $G({\bs r})$, which have been derived in doubly-asymptotic regions, using the above asymptotic formulae for the correlation length, and then shown numerically to hold for practically all distances $|{\bs r}|$. These laws can be written in a compact form as follows:
\begin{equation}
|G_l(|{\bs r}|)| \approx \epsilon_l^{\gamma} g_l \left( \epsilon_l^{\nu_l} |{\bs r}|\right), \,\,\,
\textrm{for sufficiently small $\epsilon_l$,}
\label{G scaling}
\end{equation}
where $\epsilon_l$ measures the distance on approaching a critical point, and  $ g_l$ is a scaling function whose explicit form we do not provide.
The index $l$ labels 11 cases specified as follows: $d{=}1$-model and the type of critical point: at $J$-axis ($J$-c.p.) or two kinds of critical points at $\mu$-axis ($|\mu|{\neq} 1$ or $|\mu|{=}1$), the symmetric model and the type of critical point ($J$-c.p. or $|\mu|{\neq} 1$ or $|\mu|{=}1$ ) and the type of lattice direction ($n{=}1$ or $n{\neq}1$), the antisymmetric model and the type of lattice direction (any critical point at the lines $|\mu|{=}2$  which is not too close to $\mu$-axis is admissible).
It is worth to recall here that in every considered case, the doubly-asymptotic formula for $|G_l(|{\bs r}|)|$ consists of three factors: a distance-independent positive coefficient ${\cal{C}}$, a damping factor ${\cal{D}}$, determining decay of correlations with distance (a product of an exponential and power factors), and an oscillating factor ${\cal{O}}$, so that
$|G_l(|{\bs r}|)| \approx {\cal{C}}{\cal{D}}{\cal{O}}$. In the two, out of 11 cases, where the considered critical points are located at $\mu$-axis and do not coincide with the end critical points, the scaling law (\ref{G scaling}) applies only to the product ${\cal{C}}{\cal{D}}$. The parameter $\epsilon$ depends on the type of model and the type of critical point:
\begin{eqnarray}
\epsilon = \left\{ \begin{array}{lll}
|\mu|, & \textrm{for $d{=}1$ or S}, &  J\textrm{-c.p.,}\\
|J|, & \textrm{for $d{=}1$ or S}, & \textrm{ any c.p. at $\mu$-axis,} \\
|\mu|-2, & \textrm{for AS,} & \textrm{ any c.p. at $|\mu|{=}2$-lines.} \\
\end{array} \right.
\label{delta}
\end{eqnarray}
The exponent $\nu_l$ in (\ref{G scaling}) stands for one of the universal critical exponents $\nu$, it is defined in (\ref{numu}) and (\ref{nuJ}). Finally, the exponent $\gamma$ is defined as follows:
\begin{eqnarray}
\gamma = \left\{ \begin{array}{llll}
1, & \textrm{for $d{=}1$}, &  \textrm{$J$-c.p. or $|\mu|{\neq}1$-c.p.,} & \\
1/2, & \textrm{for $d{=}1$}, &  \textrm{$|\mu|{=}1$-c.p.,} & \\
1, & \textrm{for S,} & \textrm{$J$-c.p. or $|\mu|{=}2$-c.p.}, & \textrm{any $n$,}\\
3/2, & \textrm{for S}, &\textrm{$|\mu|{\neq}2$-c.p.}, & \textrm{any $n$,} \\
1, & \textrm{for AS}, & \textrm{any c.p. at $|\mu|{=}2$-lines,} & \textrm{any $n$.}
\end{array} \right.
\label{gamma}
\end{eqnarray}
The scaling law (\ref{G scaling}) is well illustrated in
Figs.~\ref{cf_J0001}--\ref{cf_mu01_} -- for the one-dimensional model,
in Figs.~\ref{cf_J01_ry1rx}--\ref{cf_mu1_ry1rx} -- for the symmetric model and the diagonal lattice direction,
in Figs.~\ref{cf_mu01_J01_new}--\ref{cf_mu1_ry0} -- for the symmetric model and an offdiagonal lattice direction,
in Figs.~\ref{cf_J001_ry1rx} and \ref{cf_J05_ry1rx} -- for the antisymmetric model and the diagonal lattice direction,
and in Figs.~\ref{cf_J01_ry3rx} and \ref{cf_J05_ry3rx} -- for the antisymmetric model and an offdiagonal lattice direction.

\section{\label{summ} Summary}
The set of models, where theories of quantum phase transitions can be tested and illustrated is rather limited. Typically, one-dimensional spin models, as  anisotropic XY chains and Ising chains in transverse magnetic fields, are used for such purposes.
The major purpose of this article is to propose and analyze $d$-dimensional models, $d{\geq} 1$, that satisfy the following requirements:
(i) the models exhibit continuous quantum phase transitions, (ii) analytic expressions for correlation lengths as functions of model parameters can be obtained and the values of critical indices $\nu$ in critical neighborhoods of quantum-critical points can be determined, (iii) high-precision numerical calculations of correlation functions, quantum fidelity and other  quantities of interest can be carried out not only for small systems but also for macroscopic ones (whose correlation length is  considerably smaller than the linear size of the considered system). In the paper we demonstrate that at least some models of the class of lattice fermion models, defined by Hamiltonian (\ref{ham1}), satisfy those requirements in one- and  two-dimensional cases.

The large-distance asymptotic formulae for one-body reduced density matrix (an example of two-point correlation functions) $G(|{\bs r}|)$:
(\ref{corr G asympt}) -- for the one-dimensional model,
 (\ref{G_diag_asympt_2d+}) -- for the symmetric model and the diagonal lattice direction,
(\ref{G offdiag_asympt_dist_2d+}) -- for the symmetric model and offdiagonal lattice directions,
(\ref{G diag_asympt 2d-}) -- for the antisymmetric model and the diagonal lattice direction,
and (\ref{G offdiag 2d- asympt}) -- for the antisymmetric model and  offdiagonal lattice directions,
together with analytic expressions for the corresponding correlation lengths, constitute the main results of our study. To the best of our knowledge, our results concerning direction-dependent correlation lengths and critical indices in two-dimensional models are unprecedented in physics literature.

Among interesting conclusions that can be derived from these results are the scaling laws of $G({\bs r})$ and its correlation length, in various asymptotic regimes, such as neighborhoods of all types  of critical points  and the asymptotic regime of $|J| \to \infty$ for some $\mu$ separated from zero. They are presented in formulae (\ref{xi large J}), (\ref{xi small J}), and (\ref{G scaling}).

Our knowledge of critical indices $\nu$ is summarized in the phase diagrams, Fig.~\ref{diagram1d} -- for the one-dimensional system, Fig.~\ref{diagram2d+} -- for the symmetric two-dimensional system and Fig.~\ref{diagram2d-} -- for the antisymmetric two-dimensional one.

Finally, we mention that the one-dimensional version of the proposed models has already been used by us to study scaling properties of quantum fidelity of small and macroscopic systems, including  a crossover regime \cite{ajk-2}, and comparing those scaling properties with the predictions of the scaling theory of quantum fidelity. To the best of our knowledge, similar tests of the scaling theory but in two dimensions, for many reasons more intriguing than the one-dimensional case, so far have never been reported. This is the task that we carried out recently, relying on the results presented in this paper; it will be reported in a coming publication \cite{ajk-3}.\\

\begin{center}
{\bf Acknowledgements}\\
The presented studies have been supported by the University of Wroc\l aw through the projects Nr 1354/M/IFT/13 and Nr 1009/S/IFT/13.
\end{center}

\end{document}